\documentclass[12pt]{article}
\usepackage{makeidx}
\usepackage{multirow}
\usepackage{multicol}
\usepackage[dvipsnames,svgnames,table]{xcolor}
\usepackage{graphicx}
\usepackage{epstopdf}
\usepackage{xcolor}
\usepackage{ulem}
\usepackage{hyperref}
\usepackage{amsmath}
\usepackage{amssymb}
\usepackage[toc,page]{appendix}
\author{sciences.univers@gmail.com}
\title{}
\usepackage[paperwidth=612pt,paperheight=792pt,top=70pt,right=70pt,bottom=70pt,left=70pt]{geometry}

\makeatletter
	{\par\setlength{\parindent}{#3}
	\setlength{\leftmargin}{#1}       \setlength{\rightmargin}{#2}%
	\advance\linewidth -\leftmargin       \advance\linewidth -\rightmargin%
	\advance\@totalleftmargin\leftmargin  \@setpar{{\@@par}}%
	\parshape 1\@totalleftmargin \linewidth\ignorespaces}{\par}%
\makeatother 
\DeclareUnicodeCharacter{2061}{X}

\begin{document}

\begin{center}
{\Large \textbf{Title} A comparative study of MOND and MOG theories versus
the $\kappa{}$-model: An application to galaxy clusters}
\end{center}

\begin{center}
{\Large G. Pascoli }
\end{center}

\begin{center}
{\Large Email: \href{mailto:pascoli@u-picardie.fr}{pascoli@u-picardie.fr}}
\end{center}

\begin{center}
{\Large Facult\'{e} des sciences}
\end{center}

\begin{center}
{\Large D\'{e}partement de physique}
\end{center}

\begin{center}
{\Large Universit\'{e} de Picardie Jules Verne (UPJV)}
\end{center}

\begin{center}
{\Large 33 Rue Saint Leu,  Amiens, France}
\end{center}

\Large{\textbf{Abstract}}
Many   models have been  proposed to minimize the
dark matter (DM) content in various astronomical  objects at every scale in the Universe.  The most widely 
known model is MOdified Newtonian Dynamics (MOND). MOND was first published by Mordehai
Milgrom in 1983 (Milgrom, 1983; 2015; see also Banik and Zhao, 2022 for a
review). A second concurrent model is modified gravity (MOG), which is a covariant
scalar-tensor-vector (STVG) extension of  general relativity (Moffat, 2006;
2020). Other theories  also exist  but have not been  broadly applied to a large list of astronomical objects (Mannheim and 
Kazanas, 1989; Capozziello and De Laurentis, 2012; O'Brien and Moss, 2015;
Verlinde, 2017). Eventually, we can also mention  the   Newtonian Fractional-Dimension Gravity (NFDG)
(Varieschi,  2020, 2023), a gravity theory  based on  spaces with fractional (i.e., non-integer) dimension (Varieschi, and Calcagni, 2022; Calcagni, 2012).

 A new model, called $\kappa$-model,  based on very
elementary  phenomenological  considerations,  has recently  been
proposed  in the astrophysics  field. This model shows that  the  presence of dark matter can be considerably minimized  with regard to 
the dynamics of galaxies  (Pascoli, 2022 a,b). The $\kappa{}$-model  belongs to
the general family of theories descended from MOND. Under this family of theories, there is
no need to  develop a highly uncertain dark matter sector of physics to
explain the observations.

\vspace{10pt}
{\raggedright
\textbf{Keywords}: dark matter,  MOND, modified gravity,  $\kappa{}$-model,    galaxies,
 galaxy clusters
}

\section{Introduction}

The dark matter (DM) paradigm  is considered to simply explain the dynamics in individual galaxies and galaxy clusters. Astrophysicists have been searching for DM evidence for years. However, the
quantity of dark matter required is immense, and  the ratio of  the dark
matter (DM) to the baryonic component (B), (DM/B),  could largely exceed  10 in a few
galaxy clusters, raising serious doubts  about the validity of this hypothesis. A rapid  explanation to
remedy this  problem  is to say that a large quantity of invisible baryonic
matter   is not counted  in the galaxies and galaxy clusters. Unfortunately, this immediate
solution is may be  acceptable up to a factor of  2, but fully irrealistic up
to a  factor of 10. What is then the nature of dark matter ? To
date, no particle of dark matter has never been detected in lab
(DAMA/LIBRA Collaboration, 2022), even though researchers have built larger and
more sensitive detectors (Xenon Collaboration, 2023). This is rather an
intriguing status and the  DM paradigm  sounds suspiciously similar to the phlogistic and the
aether  in the nineteenth  century. Additionally, some agnostic physicists believe
that DM does not exist at all and have instead proposed alternative models of
gravity.  As a result, all sorts of theories have been built  in order to remove dark
matter in all astrophysical systems. There are two theories that stand out because they  have
been applied to numerous  concrete situations. The first is modified
Newtonian dynamics (MOND). MOND diverges from the standard Newton's laws at extremely low
accelerations, which are a characteristic of the outer regions of galaxies
(Milgrom, 1983, 2015). MOND postulates a modification to Newton's second
law, such that the
force  applied on a particle  is no  longer  proportional to the 
acceleration $a$,  but rather to its square  $a^2$, when the accelerations  are smaller than the critical limit $a_0={10}^{-10} \ m s^{-2}$ (Milgrom,
1983). This model effectively explains the dynamics of individual galaxies without dark
matter (Famey and Mc Gaugh, 2012).

 In modified gravity (MOG) (STVG or scalar-tensor-vector gravity), the approach is
very different. The structure of space-time is described by the usual  metric
tensor $g_{\mu{}\nu{}}$,  complemented by a vector field, ${\phi{}}_{\mu{}}$, and two scalar
fields, $G$ and $\mu{}$, which represent a dynamical version of
the Newtonian gravitational constant and the mass of the vector field,
respectively (Moffat, 2006; 2020). The vector field part  produces a Yukawa-like modification of
the gravitational force
due to a point source. This model accurately explains not only the
dynamics of galaxies  but also the dynamics in  galaxy clusters without dark
matter (Moffat, 2020).

The common and main  objective  of MOND and MOG theories is to eliminate in whole or
in part  the dark matter in the Universe.   The agreement between  these two
models and the observational data are very remarkable for galactic dynamics; however, the situation is  distinct for galaxy clusters where MOG appears  to have an advantage over its competitor  MOND (Brownstein and Moffat,
2006). MOG achieves to eliminate all dark matter content in these
objects, but MOND still needs to consider some form of invisible matter in galaxy clusters (Mc Gaugh, 2015).
However, this does not mean that MOND has been falsified,  because
the MONDian world is very rich and there are numerous extensions, such as
the promising  extended MOND (EMOND) (Hodson and Zhao, 2017a, b)\footnote{The
most complete  theory regarding  the   MONDian  world  is that of Solznik and
Skordis (2021), but the formalism  EMOND is much easier to manipulate  (the
theory of Solznik and Skordis belongs to the large class  of TeVeS
theories, such as MOG with many free parameters and  some
 disguised aspect of DM).}. Moreover, we can speculate whether MOG does not reintroduce  DM
in a disguised manner (the vector field ${\phi{}}_{\mu{}}$  is massive). 
Following this simple statement, only  MOND is truly free of DM,
and the path followed by  EMOND is potentially preferable.

The third model studied here is the $\kappa{}$-model; it is based on a
phenomenological and MONDian  procedure whose main  aim is to
simultaneously  explain the dynamics of  individual galaxies by minimizing the
dark matter content and  by  maintaining the  formal aspect of the  Newtonian law of
gravitation. It is based on a relational consideration; the mean density of
matter is estimated  at a very large scale, and the surroundings of  a given observer
influence the measurements made for the determination of the velocities and accelerations. In this regard, the
observations depend not only on the reference frame of
the observer but also on his environment. Thus, the $\kappa{}$-model uses a
holistic or Machian approach. The velocities and accelerations that are
environment-dependent are renormalized, and the observer measures apparent
quantities depending on a coefficient denoted $\kappa{}$. An empirical (and universal)
 relationship is provided between the coefficient $\kappa{}$ and the mean
density. The  coefficient  $\kappa{}$ intervenes in the acceleration 
term of the dynamics equation and imparts a MOND-type appearance to the $\kappa{}$-model.
  However, a  physical support is now
provided; this is the  environment of the observer, which distorts  the
measurements. Thus, a naive  analogy is that of an observer placed in a medium of  given
refraction index, who sees a magnification  of both the size and  the  velocity of any object.
However, this very simplistic comparison is not to be taken at face value because in a medium of given refraction index, the light is attenuated and  the
  images can be extremely blurred. On the contrary, in the $\kappa{}$-model framework,  no such
medium is existing;     the light   is not attenuated  and it still
propagates in straight line with the  speed $c$,  constant
and  independent of frequency, as measured in vacuum by every  observer (Pascoli, 2022a)\footnote{Further  information is summarized in the two appendices A and B placed at the end of this paper. Initially, we start with an isotropic  and homogeneous base  space $\Sigma$. In this space the light propagates in straight line. However, any observer is located at the center of a homogeneous and isotropic  universe $\kappa\Sigma$,  homothetic to the base space $\Sigma$; the coefficient $\kappa$ is dependent on the environment of this observer (the mean density of matter surrounding the observer). Unfortunately, it is very difficult to have  a global view of the situation  with  a unique $\mathbb{R}^3$-type space. Rather, the image needs to be that of a fiber bundle $\Sigma \times \kappa$  with $\Sigma$ as the base and $\kappa$ as the fiber (specifically, each observer projects all structures present in the Universe on his own stratum labelled by $\kappa$). Within  the framework of  this fiber bundle,  DM effects can be re-interpreted as a $\kappa$-lensing, whose  the Bullet Cluster is an illustrative example (paragraph 3).}.  The great
interest of  the $\kappa{}$-model is that  no arbitrary parameter is introduced,
contrary to MOG, which has two outer free parameters 
(these two models having a universal relationship for the fits).
Another  advantage  of the $\kappa{}$-model is that it would permit to
pass directly for any type of galaxies from the data on the spectroscopic velocity measurements
to the mean densities (and inversely) without the ambiguous transition through  the mass-to-light ratio
knowledge. The relationship between the spectroscopic velocities   and  the mean densities is direct in the $\kappa{}$-model. Strongly contrasting with this  view,  DM  can fit almost all 
rotational curves with any mean baryonic density profiles due to its very  large  flexibility, and for this reason the DM paradigm
  has unfortunately  no predictive value. Eventually the $\kappa{}$-model
can easily  help  to understand the weaker "DM effect" in the regions where the mean density is high (globular clusters, core
of the galaxies)  and, conversely,  the stronger "DM effect" in the regions where the mean
density is weaker (outer regions of the galaxies, low brightness galaxies and
galaxy clusters). More specifically, the $\kappa$-model predicts that when the mean density $\bar{\rho}$ in a  large-scale object (galaxy or galaxy cluster) is smaller than a critical value  $\sim 4\ 10^{-24}\  g\  cm^{-3}$\footnote{This value corresponds to  the mean density of the baryonic matter detected in the solar neighborhood (or, correspondingly,  $\sim 70\ M_\odot\  pc^{-2}$,  assuming a vertical thickness $\sim 1\ kpc$  (Famaey, and McGaugh, Fig.19, 2012)).}  the  measured velocities appear  magnified compared to the estimated newtonian velocities.

The $\kappa$-model has been already applied to various types of galaxies, such as large or small low surface brightness galaxies (LSBs) and high surface brightness galaxies (HSBs)   (Pascoli, 2022 a,b). The $\kappa$-model fits and the observational curves has been shown in fairly good agreement. Even though  this study  needs to be extended to a larger database, such as SPARC   (Lelli,
  McGaugh and  Schombert; 2017), these initial results and deductions provided by  the  $\kappa$-model are highly  encouraging.

While the $\kappa{}$-model appears to  succeed at explaining  the dynamics of numerous
individual galaxies (Pascoli, 2022), to  date, this model has not been
tested  on the galaxy cluster field, and this is  the goal of our present 
study. We show that the  $\kappa{}$-model   can greatly  lower  the  major
part of the dark matter content in the galaxy clusters. The current  ratio DM/B
amounts around $10$ in  the outskirts of these objects,  and the $\kappa{}$-model
can reduce  this ratio to approximately $0-1$. The situation is  more difficult in the inner regions of the galaxy clusters,  but  by lowering the gas temperature in these regions, the problem can be easily solved.   The $\kappa{}$-model provides a good predication for the
ratio DM/B in the outer regions of galaxy clusters, without introduction of  numerous free parameters other than those  linked to the mean density  of baryonic
matter  (Pascoli, 2022 a,b).  Likely, this result is not a mere coincidence, and
supports the consideration of our new proposal. Here,   the $\kappa{}$-model is applied
to a sample of galaxy clusters (paragraph 2) and then, eventually, to the
Bullet Cluster (paragraph 3).

\section{Quick review of the $\kappa$-model}

In this section we summarize the main ideas  developed in the preceding  papers (Pascoli and Pernas, 2020;
Pascoli, 2022 a,  b) as for the motion of an individual particle seen in a frame of reference. The appendix  A  specifies  what we mean  by  frame of reference  in the approach  of the $\kappa$-model.

In a current  manner for a terrestrial  observer $E$   the motion of any 
particle of mass  $m$   is simply 
determined  by the following dynamic equation:

\begin{equation}
m \frac{d{(\kappa{}}_E\dot{\boldsymbol{\sigma{}}})}{dt}=\boldsymbol{f}_E
\end{equation}

 where $\dot{\boldsymbol{\sigma{}}}$  is the bare velocity  of the particle (the point on $\boldsymbol{\sigma{}}$ designates the derivative of the bare position of the particle
given by  $\boldsymbol{\sigma{}}$) and $\boldsymbol{f}_E$ is the sum of forces applied to it\footnote{ The force $\boldsymbol{f}$ is the resultant force of the gravitational, electrostatic and magnetic forces, and also of the centrifugal and Coriolis forces if the frame of reference is not
inertial.}, as
evaluated  by the   observer $E$ (the coefficient $\kappa$ is defined below).   Additionally, the $\kappa{}$-model
assumes that  at a very large (galactic) scale, the immediate environment of a particle plays a
role in the determination of its  motion. For instance, the particle "feels" the
presence of the other particles in such a way that  its velocity  is   modified
by a scale factor  $\kappa{}$  that is proportional to the mean density $\bar{\rho{}}$; more exactly, this factor is
proportional  to the logarithm of the mean density. In a more
concrete manner, all structures of the Universe are usually
inserted  in a homogeneous and isotropic base space $\Sigma{}$ with coordinates $\boldsymbol{\sigma{}}$.  A
phase space  $\Pi{}(\boldsymbol{\sigma{}}, \dot{\boldsymbol{\sigma{}}})$   is attached to $\Sigma{}$ (the
point on the letter  designates the  time
derivative).  However, every   observer $O$ has his own measuring
gauge, which is dependent on his  environment. Thus, this  observer   applies   an apparent 
isotropic and homogeneous  dilation  to  the phase space  $\Pi{}(\boldsymbol{\sigma{}}, \dot{\boldsymbol{\sigma{}}})$\footnote{ Once this operation is achieved, the quantity 
$\kappa{}\boldsymbol{\sigma{}}$ forms an inseparable unit   $\boldsymbol{R}$, where $\kappa{}$ and 
$\boldsymbol{\sigma{}}$ are no longer  separately measurable for any observer  (similarly for 
$\kappa{}\dot{\boldsymbol{{\sigma{}}}}$, which forms an inseparable unit   $\dot{\boldsymbol{{R}}}$). Within the $\kappa$-model  framework the Newtonian law  is formally maintained, as follows: $m \frac{d\boldsymbol{P}}{dt}=-\boldsymbol{\nabla}_{\boldsymbol{R}}[(\Phi(\boldsymbol{R})]$ where $\boldsymbol{P}=m\dot{\boldsymbol{R}}$  is the impulsion of the particle and $\Phi$ is the gravitational potential  measured in situ.
For all  these reasons   the variation of  $\kappa{}$  is hidden to any observer, and
the space $\Sigma{}$ of vectors $\boldsymbol{{\sigma{}}}$ is not reachable.  However, the ratios of the type 
$\frac{\kappa{}'}{\kappa{}}$  are  always measurable for any observer, an
operation  which helps to  exchange the information with other observers (Pascoli, 2022).}

\begin{equation}
(\boldsymbol{\sigma{}}, \dot{\boldsymbol{\sigma{}}})\ \longrightarrow{}(\kappa{}\boldsymbol{\sigma{}}, 
\kappa{}\dot{\boldsymbol{\sigma{}}})
\end{equation}

where $\kappa{}$ is a renormalization coefficient (for the terrestrial observer $\kappa{}$ is denoted ${\kappa{}}_E$). For any  observer $O$, the coefficient  $\kappa{}$ is considered to be  independent of the point, and any other   observer  $O’$, placed in a distinct  environment, follows the  same reasoning,  but with the essential  difference that  this one  applies a coefficient  $\kappa{}'\neq\kappa{}$.  The basic idea is that any observer does not directly
access to the  phase space  $\Pi{}$ but rather  visualizes  an apparent  homothetic  replica  $\kappa{}\Pi{}$. The
$\kappa{}$-model is reduced to this sole operation.  Consequently,
this model is not a "theory" by itself.  An underlying  theory is needed, such as the 
Newtonian mechanics or the  general relativity.

For any extended objects (a galaxy or a galaxy  cluster) with a mean density profile  with  both definite  upper  ${\rho{}}_M$
and lower ${\rho{}}_m$ bounds, such as  ${\rho{}}_M > {\rho{}}_m$, there exists a
universal relationship for $\kappa{}$; this
factor is linked to the local mean density $\bar{\rho{}}$, as follows:

\begin{equation}
\frac{{\kappa{}}_M}{\kappa{}}=1+Ln (\frac{{\rho{}}_M{-\rho{}}_m}{\bar{\rho{}}-{\rho{}}_m})
\end{equation}

For a galaxy or a cluster of galaxies ${\rho{}}_M\gg{}{\rho{}}_m\simeq{}0$, this relationship
simplifies in the form  $\frac{{\kappa{}}_M}{\kappa{}}=1+Ln\ (\frac{{\rho{}}_M}{\bar{\rho{}}})$  (the function $\bar{\rho{}}$  has a
definite
upper  bound  ${\rho{}}_M$ and ${\rho{}}_m\simeq{}0$)\footnote{  However, in order to 
perform  the analysis of the CMB in the framework of the $\kappa{}-$model,  the
complete form  has to be used because ${\rho{}}_M{\simeq{}\rho{}}_m$ 
($\frac{{\rho{}}_M{-\rho{}}_{m\
}}{{\rho{}}_m}\sim{}{10}^{-5}$).}. The quantity ${\rho{}}_M$  is a reference
value  for the mean density. For any local observer the density
measured, ${\bar{\rho{}}}_{l}$, is equal to the following:

\begin{equation}
\frac{\bar{\rho{}}}{{\left[1+Ln \left(\frac{{\rho{}}_M}{\bar{\rho{}}}\right)\right]}^3}
\end{equation}

The density ${\bar{\rho{}}}_{l}$,  measured in situ, can be considered as real, whereas the density $\bar{\rho{}}$  measured by a terrestrial observer  is  only an "apparent" quantity. 
Eventually, the  mean density around a point, labelled by   $\boldsymbol{\sigma{}}$, can be obtained
by integrating over a suitably sized volume $\omega{}$ surrounding this point as follows:

\begin{equation}
\bar{\rho{}}(\boldsymbol{\sigma{})}=\int_{\omega{}}d^3{\boldsymbol{\sigma{}}}'w(\boldsymbol{\sigma{}}-\boldsymbol{\sigma{}}')\rho{(\boldsymbol{\sigma{}}')}
\end{equation}

For simplicity, we  can assume that the  spread function $w\left(\boldsymbol{\sigma{}}\right)$ is isotropic and Gaussian, i.e. $w\left(\boldsymbol{\sigma{}}\right)=(\pi\delta^2)^{(-3/2)}e^{-\left(\frac{{\boldsymbol{\sigma{}}}^2}{{\delta{}}^2}\right)}$. This  spatial averaging is used to smooth the strongly varying density distribution of matter in a galaxy over distances of a few parsecs around each point.

\vspace{20pt}

By using the transform (2), the dynamic equation (1)  can now be rewritten as follows:\footnote{ The equation  (1) is evidently not correct, and the
well known  consequence of its misuse  constitutes the missing mass problem.
The solution acknowledged today by a common consensus is to add dark matter.  Thus, we have the following relationship:

\begin{equation}
m\frac{d}{dt}\left({\kappa{}}_E\dot{\boldsymbol{{\sigma{}}}}\right)={\boldsymbol{f}}_E+\boldsymbol{f}_{DM}=(1+\alpha{}){\boldsymbol{f}}_E
\end{equation}

The observations provides $\alpha{}\sim{}1-10$, following the type of extended 
objects under consideration (the coefficient $\alpha{}$, or more generally
 the profile of the function $\alpha{}$ when $\alpha{}$ varies in the
objects,  is predominantly  adapted in an ad hoc manner).  For instance, $\alpha{}$ is small in  dense systems, such as  globular  clusters and, at the opposite side,  high
in diffuse low-density  galaxies.  
}

\begin{equation}
m\frac{d}{dt}\left(\kappa{}\dot{\boldsymbol{\sigma{}}}\right)={\left(\frac{{\kappa{}}_E}{\kappa{}}\right)}^2{\boldsymbol{f}}_E=\boldsymbol{f}
\end{equation}

where the second equality is consistent with the following relationship:

\begin{equation}
\boldsymbol{f}_{C\longrightarrow{}P}=-GM m\frac{\kappa{}
\bold{CP}}{{\kappa{}}^3{{CP}}^3}={\left(\frac{{\kappa{}}_E}{\kappa{}}\right)}^2\boldsymbol{f}_{E,C\longrightarrow{}P\
}
\end{equation}

This relationship applies for a test particle $P$ of mass $m$ under the action of a central attractor  $C$ of mass $M$ $(\boldsymbol{\sigma{}}=\boldsymbol{CP})$. The real  force is evaluated where  the particle $P$ resides (the real force is measured in situ, contrarily to the  force estimated  by the terrstrial observer which is apparent).

For a  particle  placed on a circular orbit  and subjected to a central attractor of mass $M$, the coefficient $\kappa$ is quasi-constant, and the following relationship is obtained for the spectroscopic velocities:

\begin{equation}
{\boldsymbol{v}}_{spectro}^2=(\kappa \dot{\sigma})^2=\frac{\kappa_E}{\kappa}(\frac{GM}{\kappa_E \sigma})
\end{equation}

 Introducing the apparent  radial coordinate $r=\kappa_E \sigma$, we find

\begin{equation}
{\boldsymbol{v}}_{spectro}^2=(\frac{\kappa_E}{\kappa})(\frac{GM}{r})
\end{equation}

Two very distinct interpretations of the magnification factor $\frac{\kappa_E}{\kappa}$ are then  possible: an apparent  magnification of the gravitational constant $G$  (as in MOG) or an apparent magnification of the attractive mass $M$ (as in DM with the identification $1+\alpha=\frac{\kappa_E}{\kappa}$).
  Note that  ${\boldsymbol{v}}_{spectro}^2$  is a real quantity, i.e. observer-independent (we can replace $\kappa_E$ with any value); in other words any observer measures  the same spectroscopic velocity as that valuated by  an inertial  observer, placed in situ (i.e. a local observer located  where the  emiter of the radiation resides).

We can choose the   simple example  of an archetypical  spiral galaxy where the density is seen to vary following an exponential law as a function of $r$ in the disk. Then by using the relation (3), we find that   the ratio  $\frac{{\kappa{}}_E}{\kappa{}}$ varies as $r$. In this case  the relation  (9) naturally leads to a constant value for  the spectroscopic velocity  in the outskirts of this  galaxy (Pascoli, 2022).  In the $\kappa$-model the exponential  profile for the density in the disk naturally  implies the flatness of the rotation curves of the spiral galaxies. On the contrary in the DM model as the surface density of baryon matter  declines in the outer regions of the disk, that of the dark matter must increase
in an adhoc manner to   reproduce  the flatness.

\vspace{20pt}

Let us eventually  notice  that the renormalization of lengths (eq. 2), hypothesized  in   the $\kappa$-model,  has very likely  a link,  with  the  MOND    paradigm, where the modification of the inertia term is a kind of renormalization of the acceleration. For instance Zhao and Famaey (2012) have  suggested that    a  rescaling of the  MOND acceleration constant   $a0$ would account
for the exact spatial distribution of the residual missing mass in MOND clusters.

A link   seems to be  also existing  between the $\kappa$-model and    the Newtonian Fractional-Dimension Gravity (NFDG) (Varieschi, 2020). NFDG belongs to the large family of models of gravitational
and matter (Calcagni, 2012). Especially  NFDG is based   on a  variable  dimension function  considered as
a field $D(r)$ depending on the radial coordinate $r$. This   leads
to a gravitational potential of the form $\sim \frac{1}{r^{(D-2)}}$, i.e. $\rightarrow  \frac{1}{r}$, for $D \rightarrow 3$ (central bulky region of a galaxy) and  $\rightarrow Ln r$ for $D \rightarrow 2$ (disk),  yielding a  flat
galactic rotation curves in the outer regions of spiral galaxies.  NFDG is equipped with an appropriate  scale length $l_0$, and then  the radial coordinate $r$ is considered to be a rescaled dimensionless coordinate,
i.e. $r  \frac{1}{l_0}$.  The scale length
 $l_0$ in NFDG is  related to  $a_0$ in MOND by $a_0 \approx GM/l_0^2$ for a galaxy of mass $M$.

 Thus following all these MOND-like scenarios   the missing mass problem  could  eventually be solved  by  scale transforms depending on the size of the objects under consideration (the central region or the outskirts of a galaxy, a galaxy cluster, etc). This strong  proposal  is fundamentally different  from the DM paradigm which assumes  that $85 \%$ of the matter composing  the Universe belongs to a dark sector of the physics which is  totally  unknown to us.

\section{ Galaxy clusters}

The gas density profile for a galactic cluster can be  approximately fitted  by
the following  function (Cavaliere, and  Fusco-Femiano, 1976):

\vspace{-5pt}
\begin{equation}
\rho{}\left(r\right)={\rho{}}_M{\left[1+{\left(\frac{r}{r_c}\right)}^2\right]}^{-\frac{3\beta{}}{2}}
\end{equation}

where $\rho{}\left(r\right)$ is the intracluster medium (ICM) mass density profile and ${
\rho{}}_M$ is the  maximal value taken by $\rho{}\left(r\right)$.  $r_c$ and
$\beta{}$ are  fit parameters for\ the distribution of the density.

Due to this isotropic density distribution, the gas mass contained in a sphere of radius $r$ is as follows:

\begin{equation}
{M}_{gas}(r)= 4\pi \int_0^r dr {r'}^2\rho{}\left(r'\right) ={}_2{F_1}(1.5,\ 1.5\beta{},2.5,-0.000017{\ r}^2)
\end{equation}

where $_2{F_1}$  is the hypergeometric function.

 When  $r \gg r_c$ and $\beta<1$, we can approximate this formula by the more manipulable  relationship, as follows:

\begin{equation}
{M}_{gas}(r)=\frac{4 \pi \rho_M r_c^3}{3(1-\beta{})}(\frac{r}{r_c})^{{3(1-\beta{})}}
\end{equation}

Unfortunately this relationship is clearly divergent when $r \longrightarrow \infty$. A cut-off for the distribution of gas, necessary for finite spatial  extent,   needs to be introduced. Following Brownstein and Moffat (2006), this cut-off, let rout, is chosen equal to the radius at which the density, drops  to $10^{-28}\ g\ cm^{-3}$ or 250 times the mean cosmological density of the baryons.  

\noindent On the other hand, assuming that the
cluster is in hydrostatic (isothermal) equilibrium,  the Newtonian dynamical mass is as follows (Brownstein and Moffat, 2006, eq. 19):

\begin{equation}
M_N\left(r\right)=\frac{3\beta{}\
k_B T}{{\mu{}m}_pG}\left(\frac{r^3}{r^2+r_c^2}\right)
\end{equation}

where $T$ is the temperature, $k_B$ is Boltzmann's constant, $\mu{}\ \approx{}\ 0.609$
is the mean atomic weight and $m_p$ is the proton mass.

In eq.4 the quantities  $\beta{}$, $k_B$, $T$, $\mu{}$, $m_p$
and $G$  are $\kappa{}$-invariant, both $r$ and $r_c^2$\footnote{The temperature $T$ is measured in situ and is  observer-independent, likewise for the dispersion velocities $\sigma_r,\sigma_\theta,\sigma_\phi$   (for instance in eq. 9 of Brownstein and Moffat, 2006 : $\sigma_r^2=  (k_B T)/\mu m_p$ is observer-independent, as is the gravitational potential  $\Phi$ which is also measured in situ).}   are transformed as $r\longrightarrow{}\frac{\kappa{}}{{\kappa{}}_E}r$. Then the  $\kappa{}$-mass
profile for a cluster
is  given by the following relationship:

\begin{equation}
M_{\kappa{}}(r)=\frac{\kappa{}}{{\kappa{}}_E}M_N(r)
\end{equation}

where  ${M}_N(r)$ is the Newtonian mass evaluated at the radius $r$. Then the mean density  $\rho{}\left(r\right)$ is inserted  
in the  relationship $(\kappa{}, \rho{})$ (see the  appendix  A eq. 18
placed at the end of the article) and this  leads to the magnification ratio, as follows:

\vspace{-10pt}
\begin{multline}
\frac{{\kappa{}}_E}{\kappa{}}=1+Ln\ \left(\frac{{\rho{}}_E}{\rho{}}\right)=1+Ln
\left(\frac{{\rho{}}_E}{{\rho{}}_M}\frac{{\rho{}}_M}{\rho{}}\right) \\  =1+Ln\
\left(\frac{{\rho{}}_E}{{\rho{}}_M}\right)+\frac{3\beta{}}{2}Ln
\left[1+{\left(\frac{r}{r_c}\right)}^2\right]
\end{multline}
\vspace{-5pt}

with the mean mass density ${\rho{}}_E$  near the Sun  estimated to $4\
{10}^{-24}\ g\ {cm}^{-3}$.  We used the  sample of galaxy clusters  from the paper of  Brownstein and Moffat (2006). The relevant properties are listed in Table 1. By convention column (6) is
the position, $r_{out}$, at which the density  drops to ${\rho{}}_{out}\simeq{}{10}^{-28} \ {g\ cm}^{-3}$, or 250 times the mean cosmological density of the baryons.

Figure  1  shows the results of our analysis  applied to  the COMA
cluster.  The $\kappa{}$-curve (shown in amber)  has approximately  the same
profile as MOND (shown in  dashed-dotted cyan) when $r>100\ kpc$, but  the $\kappa{}$-curve is much
closer to the observational curve and even merges with  the latter
one  in the outer regions of the cluster. Even though the $\kappa{}$-curve is not fully
merged with the ICM gas curve inside the inner regions; however,  the apparent
gravitational mass has been largely  lowered (by a factor in the range $7-10$ along the curve). Finally a residual  gap has to be filled in between the $\kappa{}$-curve and
the observational one in the inner regions of the COMA cluster. To do this,   we
propose to decrease the temperature in the inner regions. This proposal can be
applied in the same way  to the other galaxy clusters (Figure 2). Notably, in  most of the cases  presented on this array of figures,  MOG theory  leads
exactly to the same prediction as the temperature  profiles  after comparing
the curves (we can refer to A0085, A0133, NGC 507 and A062 as examples).  Assuming a non-isothermal temperature 
profile, the Newtonian mass has to be re-calculated as follows (Brownstein and Moffat, 2006, eq. 18):

\begin{equation}
M_N\left(r\right)=\frac{r\
k_BT(r)  }{{\mu{}m}_pG}\left(\frac{3\beta{}r^2}{r^2+r_c^2}-\frac{dLnT(r)}{dLn(r)}\right)
\end{equation}

{\raggedright
Then we used  the following   easy-to-manipulate temperature profile:
}

\begin{equation}
T\left(r\right)=T_{out}exp [-\alpha{}(\frac{r_{out}-r}{r_{out}})]
\end{equation}

where $T_{out}$ designates the temperature in the outer regions of the cluster. This parameter is provided in column (2) of Table 1.  The coefficient  $\alpha{}=1$ is used when   $M_N\lesssim{}{10\ M}_{gas}$. This is the case for most
situations  under study. For a few cases where this ratio  is far beyond 10,  
then $\alpha{}=2$ is used.  The  temperature profile (eq.8) has been selected  such
that the outer temperature, $T_{out}$  (column (7) of Table 1)  is that provided by Brownstein and Moffat (2006), as directly resulting from observations (Reiprich and B\"{o}hringer, H. 2002).

\begin{equation}
M_N\left(r\right)=\frac{r\ k_BT(r)}{{\mu{}m}_pG}\left[\frac{3\beta{}r^2}{r^2+r_c^2}-\frac{\alpha{}r}{r_{out}}\right]
\end{equation}

\vspace{5pt}

The green curve  on the Figure 1 is the $\kappa{}$-profile with a non-isothermal temperature
profile given by eq. 8.  The  $\kappa{}$-curve associated to a non-isothermal temperature profile  becomes virtually  superimposed 
to the ICM  gas profile when $r < 500\ kpc$ (let us  note that for $r> 500\ kpc$, the $\kappa$-curve, associated to an isothermal temperature profile  and that  associated to a non-isothermal temperature profile  flank the  ICM gas profile). Thus, a prediction of the  $\kappa{}$-model  is  that the
temperature is lowered by a factor two (at $r_{out}/2$) in the inner  regions compared to that   in the  outskirts of the COMA cluster. More precisely, the most appropriate description  is that of a  non-isothermal core  for $r < 1000\ kpc$, which is itself surrounded by a quasi-isothermal shell when $1000 < r < 1954 \ kpc$. A vey similar situation is encountered  in the other clusters collected in  Table 1.  Thus,  since the $\kappa$-model has no outer parameters, an optimal fit can  be achieved  by finely adjusting the temperature profile and, in return,  can eventually predict  this   profile (whereas this refinement and prediction   are not possible within the ad hoc, and definitely  not falsifiable,   DM paradigm, which is easily  adapted to fit the ICM gas curve with  any temperature profile indicating  no predictive value).

\begin{center}
\includegraphics[height=220pt, width=250pt]{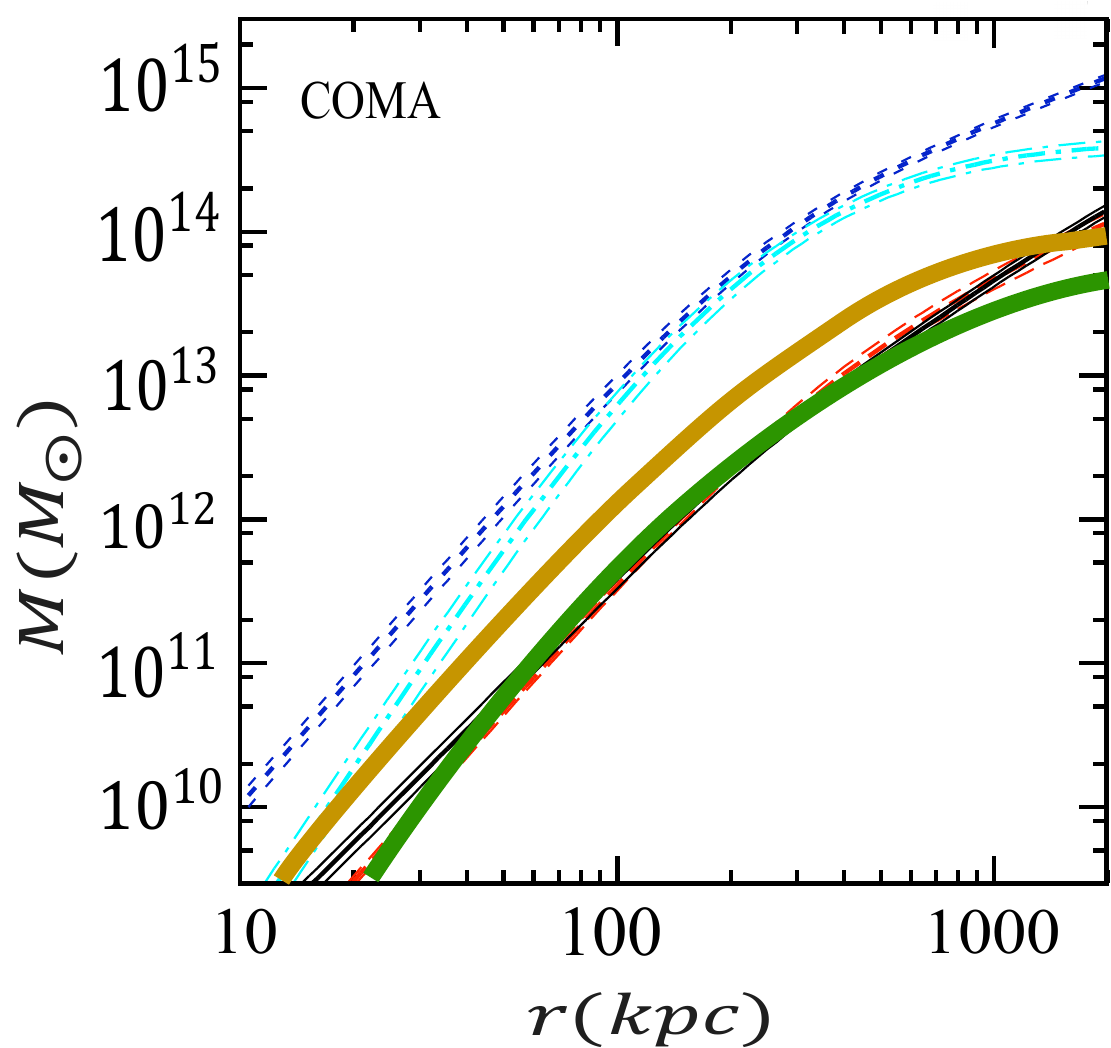}
\end{center}

Figure 1 COMA cluster profile. The horizontal axis is the radius  in kpc and the
vertical axis is mass in units of the solar mass  $M_\odot$. The red long
dashed curve is the ICM gas mass derived from X-ray observations
(compilation of Reiprich, 2001; Reiprich and B\"{o}hringer, 2002);  the short
dashed blue curve is the Newtonian dynamic mass; the dashed-dotted cyan curve is
the MOND dynamic mass; the solid black curve is the MSTG dynamic mass (Browstein and Moffat, 2006). Our contribution is displayed as  the
amber curve, showing the $\kappa{}$-model dynamic mass   with the temperature $ T=8.38\ keV$. 
The solid  green     curve displays   the $\kappa{}$-model dynamic mass, 
assuming   a non-isothermal temperature profile with $\alpha{}=1$ in eq. 8.

\vspace{15pt}

A perfect superposition with  the ICM gas mass  curve in COMA cluster can still  be achieved by taking the temperature profile displayed in  Figure 2. We note an increase of the mean  temperature  from the inner regions up to $1500\ kpc$ and then a slow decrease toward the outskirts. However, a comparison of this profile with observational data is not very conclusive.  The main reason is that  the COMA cluster modeled here, similar to other research (Browstein and Moffat, 2006), is in the form of a  spherical and well-relaxed distribution of gas. The reality is much more complex. The density in the inner  regions is relatively high, and the  cooling through thermal bremsstrahlung emission must effectively  be much more efficient   than  in the outskirts. However, the heating by active galactic  nuclei in the inner regions can compensate for the cooling processes. Thus, the energetics are very complex and the temperature profile is not  easily predicted (Bykov et al, 2015).   In a general manner  for the galaxy clusters, all temperature profiles  found in the literature are model-dependent with  a very large amount of dark matter. Moreover, even limiting to the outskirts, which are directly accessible, the physics is  poorly understood (Walker et al, 2019).   Thus, the  temperature map of the outskirts of the COMA cluster  simultaneously exhibits  cool and hot regions with various substructures (Watanabe et al, 1999).

\begin{center}
\includegraphics[height=160pt, width=210pt]{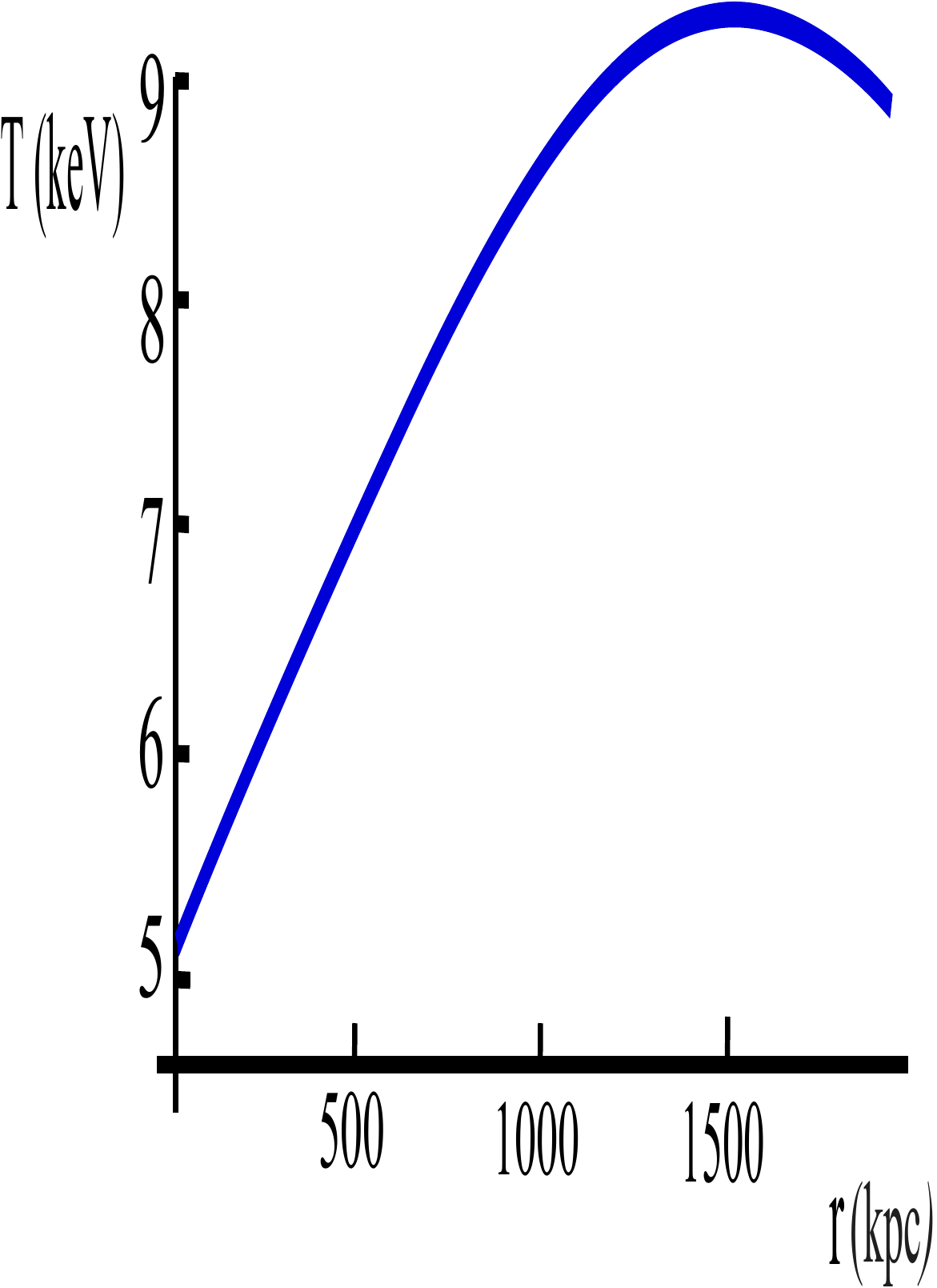}
\end{center}
Figure 2 Predicted mean temperature profile in the COMA cluster in the framework of the $\kappa$-model

 We have suggested  here  to solve the problem of the very large excess of (newtonian) mass in the galaxy clusters by using a two-stage procedure. First, by significantly reducing   the apparent attractive mass with the $\kappa$-model. With help of this operation the ratio $M_{DM}/M_{gas}$ is reduced to  rather more credible values, i.e.  $\sim 0$ in the outer regions and  to $\sim 2$ in the inner regions. Then the second step consists in  adapting  the temperature profile with  cancellation of the residual excess of mass. It seems that MOG  was facing  the same situation, even though Brownstein and Moffat (2006) have not tried to perform this second step (for instance for  A0085, ...). Another issue for this secong step would still  be to follow the MOND galaxy cluster analysis of Banik and Zhao (2022), that is to say to add a dense core composed of sterile neutrinos (see Giunti and  Lasserre (2019) for a review on these hypothetical  particles).  The  observational data not being  perfect  it is very likely that a two-stage solution is also needed in the framework of  other theories (for instance MOND) which have been proposed to eliminate the dark mater (taken into account of the measurement uncertainties of  inclination, thickness, mass-to-light ratio for the individual spiral galaxies; density/temperature profiles and clumpiness in the case of galaxy clusters, and so on). The only model which runs directly  in one step is dark matter given its undue flexibility (compared to the $\kappa$-model with no flexibility).

\begin{center}
\includegraphics[height=460pt, width=310pt]{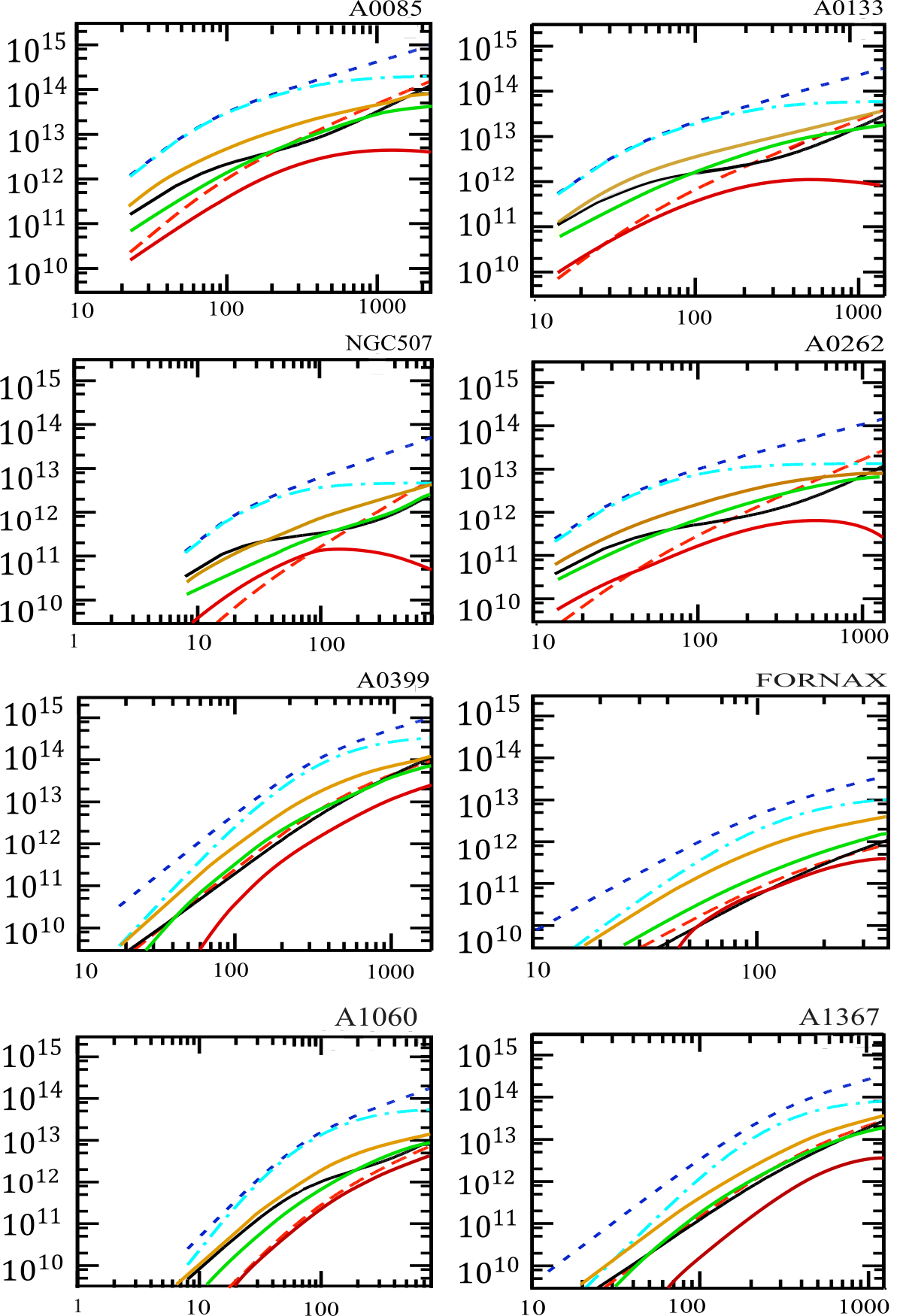}
\end{center}

Figure 3 Plot of the radial mass profile
for  clusters of the sample in Table 1. The horizontal axis expresses the radius in $kpc$ and the vertical axis is mass in units of solar mass $M_\odot$.  The red long
dashed curve is the ICM gas mass derived from X-ray observations
(compilation of Reiprich, 2001; Reiprich and B\"{o}hringer, 2002);  the short
dashed blue curve is the Newtonian dynamic mass; the dashed-dotted cyan curve is
the MOND dynamic mass; the solid black curve is the MSTG dynamic mass (Browstein and Moffat, 2006). Our contribution is displayed as  the
amber curve, showing the $\kappa{}$-model dynamic mass   with the temperature $ T=8.38\ keV$. 
The solid  green     curve displays   the $\kappa{}$-model dynamic mass, 
assuming   a non-isothermal temperature profile with $\alpha{}=1$ in eq. 8.  The solid  red    curve displays   the $\kappa{}$-model dynamic mass,
with   a non-isothermal temperature profile of $\alpha{}=2$.

\begin{center}
\includegraphics[height=440pt, width=310pt]{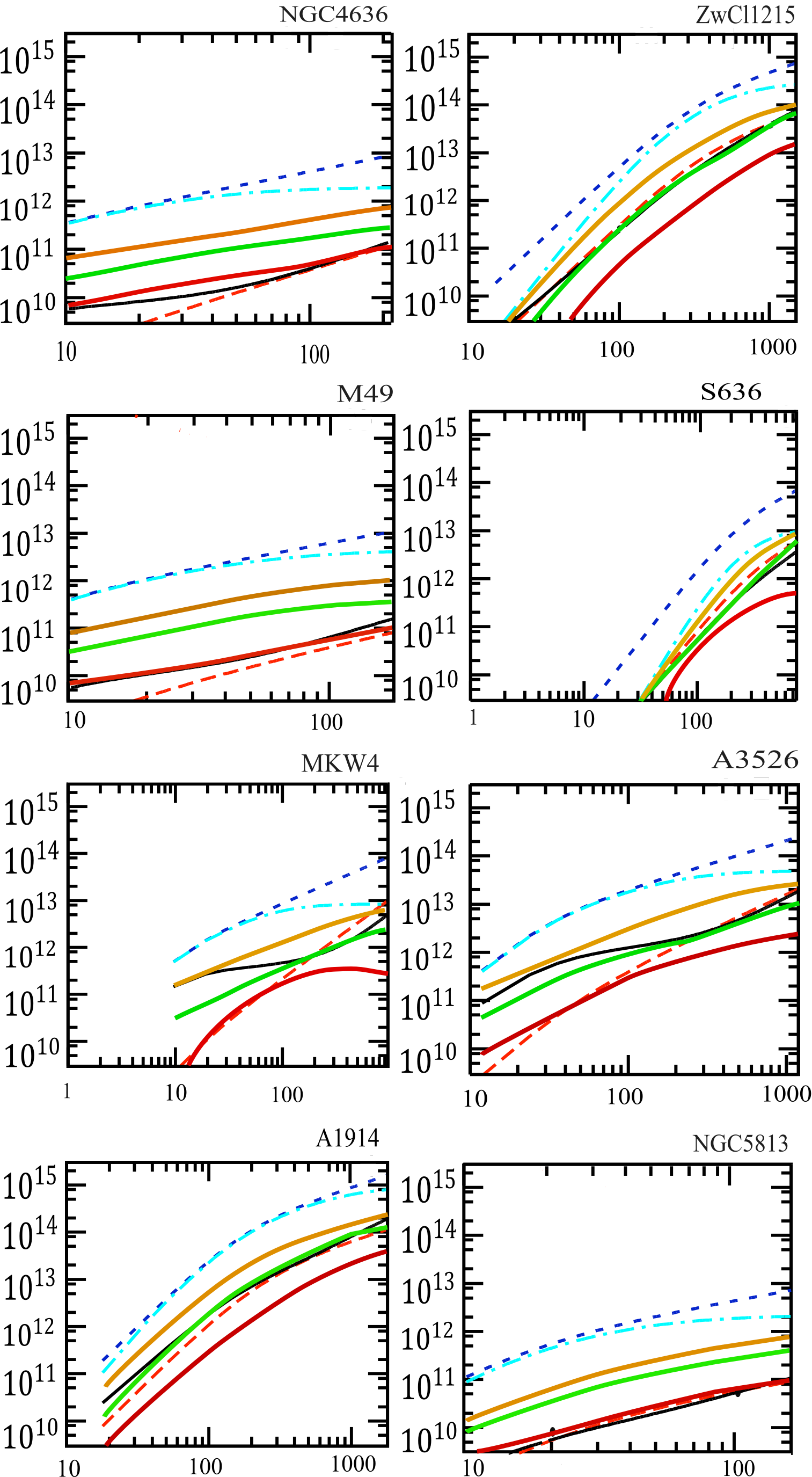}

Figure 3 Continued galaxy cluster mass profiles

\includegraphics[height=230pt, width=310pt]{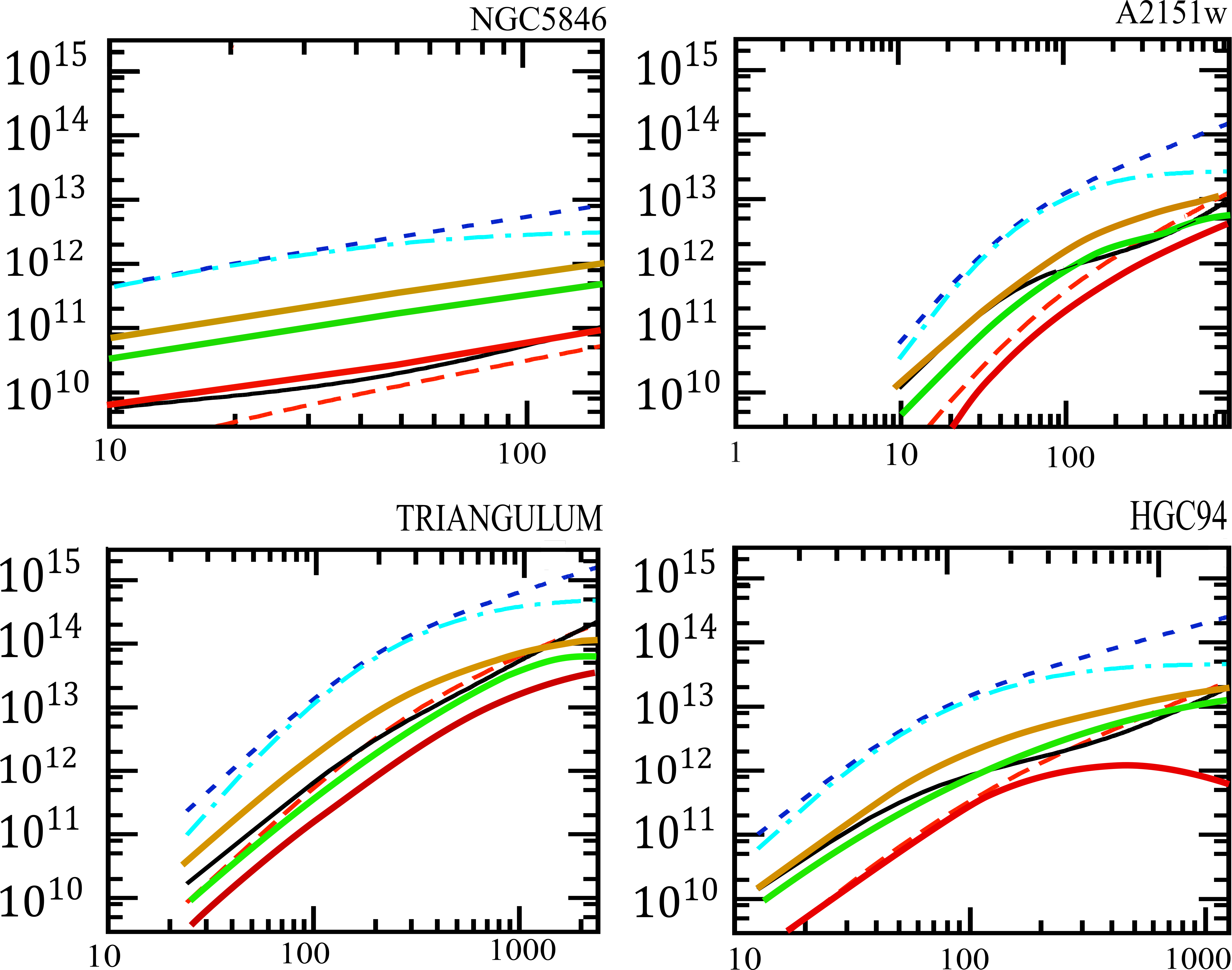}

Figure 3 Continued galaxy cluster mass profiles

\end{center}

{\raggedright
Table  1 Galaxy cluster properties
}

{\raggedright
{\tiny Note - This  compilation  is  issued  from Brownstein and Moffat (2006).
We have added a column for the  mass $M_{\kappa{}}$.}
}

{\raggedright
{\tiny Column (1) Galaxy cluster name 
 \   \  \   \   \  \   \   \  \   \   \  \  \   \        Column (6) radius where gas  $\simeq{}{10}^{-28}\ g\ {cm}^{-3}$}
}

{\raggedright
{\tiny Column (2) X-ray temperature         
           \   \  \    \   \  \    \   \  \   \   \  \  \   \  \  \       Column (7) ICM gas mass integrated to $r_{out}$}
}

{\raggedright
{\tiny Column (3) ICM central mass density  \  \  \                                     
     \   \  \    \     Column (8) Newtonian dynamic mass integrated to rout}
}

{\raggedright
{\tiny Column (4) model  parameter  \  \  \
 \   \  \   \   \  \   \   \  \  \   \  \   \   \  \  Column (9) MSTG dynamic mass integrated to rout}
}

{\raggedright
{\tiny Column (5) model core radius parameter   \  \  \                                 
Column (10) convergent MOND dynamic mass}
}

{\raggedright
{\tiny Column (11) $M_{\kappa{}}$ integrated to rout}
}

\vspace{3pt} \noindent
\begin{tabular}{|p{51pt}|p{14pt}|p{56pt}|p{28pt}|p{28pt}|p{28pt}|p{28pt}|p{35pt}|p{35pt}|p{35pt}|p{35pt}|}
\hline
\parbox{51pt}{\centering    
{\scriptsize Cluster

\vspace{8pt}
(1)
}} & \parbox{14pt}{\centering
{\scriptsize $T$

$keV$ 

(2)
}} & \parbox{56pt}{\centering
{\scriptsize ${\rho{}}_M$

\vspace{4pt}
${10}^{-25}\ g\ {cm}^{-3}$

\vspace{-1pt}
(3)}

} & \parbox{28pt}{\centering
{\scriptsize $\beta{}$ 

\vspace{9pt}
(4)}
} & \parbox{28pt}{\centering
{\scriptsize $r_c$

$kpc$

\vspace{2pt}
(5)}}

 & \parbox{29pt}{\centering
{\scriptsize

$r_{out}$

$kpc$

\vspace{2pt}
(6)
}
} & \parbox{28pt}{\centering
{\scriptsize

$M_{gas}$

\vspace{2pt}

${10}^{14}M_\odot$ 

\vspace{-2pt}
(7)}} & \parbox{35pt}{\centering

{\scriptsize   $M_N$

\vspace{10pt}
(8)
}}

 & \parbox{35pt}{\centering

{\scriptsize

$M_{MSTG}$

\vspace{10pt}
(9)}}
 & \parbox{35pt}{\centering 

{\scriptsize
$M_{MOND}$

\vspace{10pt}
(10)}} & \parbox{35pt}{\centering 
{\scriptsize

$M_{\kappa{}}$

\vspace{10pt}
(11)}}

 \\
\hline

\parbox{51pt}{\raggedright 
{\scriptsize A0085}
}

& \parbox{28pt}{\raggedright 
{\scriptsize 6.90} }& \parbox{56pt}{\centering
{\scriptsize 0.34}

} & \parbox{28pt}{\raggedright 
{\scriptsize 0.532}
} & \parbox{28pt}{\raggedright 
{\scriptsize 58.5}
} & \parbox{28pt}{\raggedright 
{\scriptsize 2,241}
} & \parbox{35pt}{\raggedright 
{\scriptsize 1.48}
} & \parbox{21pt}{\raggedright 
{\scriptsize 9.02}
} & \parbox{21pt}{\raggedright 
{\scriptsize 1.15}
} & \parbox{21pt}{\raggedright 
{\scriptsize 1.83}
} & \parbox{21pt}{\raggedright 
{\scriptsize 0.77}
} \\
\hline
\parbox{51pt}{\raggedright 
{\scriptsize A0119}
} & \parbox{14pt}{\centering 
{\scriptsize 5.60}
} & \parbox{56pt}{\centering 
{\scriptsize 0.03}
} & \parbox{28pt}{\raggedright 
{\scriptsize 0.675}
} & \parbox{28pt}{\raggedright 
{\scriptsize 352.8}
} & \parbox{28pt}{\raggedright 
{\scriptsize 1728}
} & \parbox{35pt}{\raggedright 
{\scriptsize 0.73}
} & \parbox{21pt}{\raggedright 
{\scriptsize 6.88}
} & \parbox{21pt}{\raggedright 
{\scriptsize 0.73}
} & \parbox{21pt}{\raggedright 
{\scriptsize 1.76}
} & \parbox{21pt}{\raggedright 
{\scriptsize 0.60}
} \\
\hline
\parbox{51pt}{\raggedright 
{\scriptsize A0133}
} & \parbox{14pt}{\centering 
{\scriptsize 3.80}
} & \parbox{56pt}{\centering 
{\scriptsize 0.42}
} & \parbox{28pt}{\raggedright 
{\scriptsize 0.530}
} & \parbox{28pt}{\raggedright 
{\scriptsize 31.7}
} & \parbox{28pt}{\raggedright 
{\scriptsize 1417}
} & \parbox{35pt}{\raggedright 
{\scriptsize 0.37}
} & \parbox{21pt}{\raggedright 
{\scriptsize 3.13}
} & \parbox{21pt}{\raggedright 
{\scriptsize 0.28}
} & \parbox{21pt}{\raggedright 
{\scriptsize 0.55}
} & \parbox{21pt}{\raggedright 
{\scriptsize 0.27}
} \\
\hline
\parbox{51pt}{\raggedright 
{\scriptsize NGC507}
} & \parbox{14pt}{\centering 
{\scriptsize 1.26}
} & \parbox{56pt}{\centering 
{\scriptsize 0.23}
} & \parbox{28pt}{\raggedright 
{\scriptsize 0.444}
} & \parbox{28pt}{\raggedright 
{\scriptsize 13.4}
} & \parbox{28pt}{\raggedright 
{\scriptsize 783}
} & \parbox{35pt}{\raggedright 
{\scriptsize 0.05}
} & \parbox{21pt}{\raggedright 
{\scriptsize 0.48}
} & \parbox{21pt}{\raggedright 
{\scriptsize 0.02}
} & \parbox{21pt}{\raggedright 
{\scriptsize 0.04}
} & \parbox{21pt}{\raggedright 
{\scriptsize 0.04}
} \\
\hline
\parbox{51pt}{\raggedright 
{\scriptsize A0262}
} & \parbox{14pt}{\centering 
{\scriptsize 2.15}
} & \parbox{56pt}{\centering 
{\scriptsize 0.16}
} & \parbox{28pt}{\raggedright 
{\scriptsize 0.443}
} & \parbox{28pt}{\raggedright 
{\scriptsize 29.6}
} & \parbox{28pt}{\raggedright 
{\scriptsize 1334}
} & \parbox{35pt}{\raggedright 
{\scriptsize 0.26}
} & \parbox{21pt}{\raggedright 
{\scriptsize 1.39}
} & \parbox{21pt}{\raggedright 
{\scriptsize 0.11}
} & \parbox{21pt}{\raggedright 
{\scriptsize 0.13}
} & \parbox{21pt}{\raggedright 
{\scriptsize 0.12}
} \\
\hline
\parbox{51pt}{\raggedright 
{\scriptsize A0399}
} & \parbox{14pt}{\centering 
{\scriptsize 7.00}
} & \parbox{56pt}{\centering 
{\scriptsize 0.04}
} & \parbox{28pt}{\raggedright 
{\scriptsize 0.713}
} & \parbox{28pt}{\raggedright 
{\scriptsize 316.9}
} & \parbox{28pt}{\raggedright 
{\scriptsize 1791}
} & \parbox{35pt}{\raggedright 
{\scriptsize 0.90}
} & \parbox{21pt}{\raggedright 
{\scriptsize 9.51}
} & \parbox{21pt}{\raggedright 
{\scriptsize 1.07}
} & \parbox{21pt}{\raggedright 
{\scriptsize 3.07}
} & \parbox{21pt}{\raggedright 
{\scriptsize 0.82}
} \\
\hline
\parbox{51pt}{\raggedright 
{\scriptsize FORNAX}
} & \parbox{14pt}{\centering 
{\scriptsize 1.20}
} & \parbox{56pt}{\centering 
{\scriptsize 0.02}
} & \parbox{28pt}{\raggedright 
{\scriptsize 0.804}
} & \parbox{28pt}{\raggedright 
{\scriptsize 122.5}
} & \parbox{28pt}{\raggedright 
{\scriptsize 387}
} & \parbox{35pt}{\raggedright 
{\scriptsize 0.009}
} & \parbox{21pt}{\raggedright 
{\scriptsize 0.373}
} & \parbox{21pt}{\raggedright 
{\scriptsize 0.011}
} & \parbox{21pt}{\raggedright 
{\scriptsize 0.102}
} & \parbox{21pt}{\raggedright 
{\scriptsize 0.026}
} \\
\hline
\parbox{51pt}{\raggedright 
{\scriptsize NGC1550}
} & \parbox{14pt}{\centering 
{\scriptsize 1.43}
} & \parbox{56pt}{\centering 
{\scriptsize 0.15}
} & \parbox{28pt}{\raggedright 
{\scriptsize 0.554}
} & \parbox{28pt}{\raggedright 
{\scriptsize 31.7}
} & \parbox{28pt}{\raggedright 
{\scriptsize 632}
} & \parbox{35pt}{\raggedright 
{\scriptsize 0.034}
} & \parbox{21pt}{\raggedright 
{\scriptsize 0.548}
} & \parbox{21pt}{\raggedright 
{\scriptsize 0.024}
} & \parbox{21pt}{\raggedright 
{\scriptsize 0.086}
} & \parbox{21pt}{\raggedright 
{\scriptsize 0.047}
} \\
\hline
\parbox{51pt}{\raggedright 
{\scriptsize A1060}
} & \parbox{14pt}{\centering 
{\scriptsize 3.24}
} & \parbox{56pt}{\centering 
{\scriptsize 0.09}
} & \parbox{28pt}{\raggedright 
{\scriptsize 0.607}
} & \parbox{28pt}{\raggedright 
{\scriptsize 66.2}
} & \parbox{28pt}{\raggedright 
{\scriptsize 790}
} & \parbox{35pt}{\raggedright 
{\scriptsize 0.07}
} & \parbox{21pt}{\raggedright 
{\scriptsize 1.69}
} & \parbox{21pt}{\raggedright 
{\scriptsize 0.10}
} & \parbox{21pt}{\raggedright 
{\scriptsize 0.50}
} & \parbox{21pt}{\raggedright 
{\scriptsize 0.21}
} \\
\hline
\parbox{51pt}{\raggedright 
{\scriptsize A1367}
} & \parbox{14pt}{\centering 
{\scriptsize 3.55}
} & \parbox{56pt}{\centering 
{\scriptsize 0.03}
} & \parbox{28pt}{\raggedright 
{\scriptsize 0.695}
} & \parbox{28pt}{\raggedright 
{\scriptsize 269.7}
} & \parbox{28pt}{\raggedright 
{\scriptsize 1234}
} & \parbox{35pt}{\raggedright 
{\scriptsize 0.27}
} & \parbox{21pt}{\raggedright 
{\scriptsize 3.19}
} & \parbox{21pt}{\raggedright 
{\scriptsize 0.26}
} & \parbox{21pt}{\raggedright 
{\scriptsize 0.75}
} & \parbox{21pt}{\raggedright 
{\scriptsize 0.40}
} \\
\hline
\parbox{51pt}{\raggedright 
{\scriptsize MKW4}
} & \parbox{14pt}{\centering 
{\scriptsize 1.71}
} & \parbox{56pt}{\centering 
{\scriptsize 0.57}
} & \parbox{28pt}{\raggedright 
{\scriptsize 0.440}
} & \parbox{28pt}{\raggedright 
{\scriptsize 7.7}
} & \parbox{28pt}{\raggedright 
{\scriptsize 948}
} & \parbox{35pt}{\raggedright 
{\scriptsize 0.09}
} & \parbox{21pt}{\raggedright 
{\scriptsize 0.78}
} & \parbox{21pt}{\raggedright 
{\scriptsize 0.05}
} & \parbox{21pt}{\raggedright 
{\scriptsize 0.08}
} & \parbox{21pt}{\raggedright 
{\scriptsize 0.10}
} \\
\hline
\parbox{51pt}{\raggedright 
{\scriptsize ZwCl1215}
} & \parbox{14pt}{\centering 
{\scriptsize 5.68}
} & \parbox{56pt}{\centering 
{\scriptsize 0.05}
} & \parbox{28pt}{\raggedright 
{\scriptsize 0.819}
} & \parbox{28pt}{\raggedright 
{\scriptsize 303.5}
} & \parbox{28pt}{\raggedright 
{\scriptsize 1485}
} & \parbox{35pt}{\raggedright 
{\scriptsize 0.59}
} & \parbox{21pt}{\raggedright 
{\scriptsize 7.15}
} & \parbox{21pt}{\raggedright 
{\scriptsize 0.72}
} & \parbox{21pt}{\raggedright 
{\scriptsize 2.5}
} & \parbox{21pt}{\raggedright 
{\scriptsize 0.91}
} \\
\hline
\parbox{51pt}{\raggedright 
{\scriptsize NGC4636}
} & \parbox{14pt}{\centering 
{\scriptsize 0.76}
} & \parbox{56pt}{\centering 
{\scriptsize 0.33}
} & \parbox{28pt}{\raggedright 
{\scriptsize 0.491}
} & \parbox{28pt}{\raggedright 
{\scriptsize 4.2}
} & \parbox{28pt}{\raggedright 
{\scriptsize 216}
} & \parbox{35pt}{\raggedright 
{\scriptsize 0.001}
} & \parbox{21pt}{\raggedright 
{\scriptsize 0.088}
} & \parbox{21pt}{\raggedright 
{\scriptsize 0.001}
} & \parbox{21pt}{\raggedright 
{\scriptsize 0.019}
} & \parbox{21pt}{\raggedright 
{\scriptsize 0.011}
} \\
\hline
\parbox{51pt}{\raggedright 
{\scriptsize A3526}
} & \parbox{14pt}{\centering 
{\scriptsize 3.68}
} & \parbox{56pt}{\centering 
{\scriptsize 0.29}
} & \parbox{28pt}{\raggedright 
{\scriptsize 0.495}
} & \parbox{28pt}{\raggedright 
{\scriptsize 26.1}
} & \parbox{28pt}{\raggedright 
{\scriptsize 1175}
} & \parbox{35pt}{\raggedright 
{\scriptsize 0.20}
} & \parbox{21pt}{\raggedright 
{\scriptsize 2.35}
} & \parbox{21pt}{\raggedright 
{\scriptsize 0.17}
} & \parbox{21pt}{\raggedright 
{\scriptsize 0.45}
} & \parbox{21pt}{\raggedright 
{\scriptsize 0.24}
} \\
\hline
\parbox{51pt}{\raggedright 
{\scriptsize A3266}
} & \parbox{14pt}{\centering 
{\scriptsize 8.00}
} & \parbox{56pt}{\centering 
{\scriptsize 0.05}
} & \parbox{28pt}{\raggedright 
{\scriptsize 0.796}
} & \parbox{28pt}{\raggedright 
{\scriptsize 397.2}
} & \parbox{28pt}{\raggedright 
{\scriptsize 1,915}
} & \parbox{35pt}{\raggedright 
{\scriptsize 1.22}
} & \parbox{21pt}{\raggedright 
{\scriptsize 12.82}
} & \parbox{21pt}{\raggedright 
{\scriptsize 1.56}
} & \parbox{21pt}{\raggedright 
{\scriptsize 4.79}
} & \parbox{21pt}{\raggedright 
{\scriptsize 1.12}
} \\
\hline
\parbox{51pt}{\raggedright 
{\scriptsize A3395s}
} & \parbox{14pt}{\centering 
{\scriptsize 5.00}
} & \parbox{56pt}{\centering 
{\scriptsize 0.03}
} & \parbox{28pt}{\raggedright 
{\scriptsize 0.964}
} & \parbox{28pt}{\raggedright 
{\scriptsize 425.4}
} & \parbox{28pt}{\raggedright 
{\scriptsize 1223}
} & \parbox{35pt}{\raggedright 
{\scriptsize 0.32}
} & \parbox{21pt}{\raggedright 
{\scriptsize 5.77}
} & \parbox{21pt}{\raggedright 
{\scriptsize 0.49}
} & \parbox{21pt}{\raggedright 
{\scriptsize 2.34}
} & \parbox{21pt}{\raggedright 
{\scriptsize 0.50}
} \\
\hline
\parbox{51pt}{\raggedright 
{\scriptsize COMA}
} & \parbox{14pt}{\centering 
{\scriptsize 8.38}
} & \parbox{56pt}{\centering 
{\scriptsize 0.06}
} & \parbox{28pt}{\raggedright 
{\scriptsize 0.654}
} & \parbox{28pt}{\raggedright 
{\scriptsize 242.3}
} & \parbox{28pt}{\raggedright 
{\scriptsize 1954}
} & \parbox{35pt}{\raggedright 
{\scriptsize 1.13}
} & \parbox{21pt}{\raggedright 
{\scriptsize 11.57}
} & \parbox{21pt}{\raggedright 
{\scriptsize 1.38}
} & \parbox{21pt}{\raggedright 
{\scriptsize 3.81}
} & \parbox{21pt}{\raggedright 
{\scriptsize 0.99}
} \\
\hline
\parbox{51pt}{\raggedright 
{\scriptsize A2065}
} & \parbox{14pt}{\centering 
{\scriptsize 5.50}
} & \parbox{56pt}{\centering 
{\scriptsize 0.04}
} & \parbox{28pt}{\raggedright 
{\scriptsize 1.16}
} & \parbox{28pt}{\raggedright 
{\scriptsize 485.9}
} & \parbox{28pt}{\raggedright 
{\scriptsize 1302}
} & \parbox{35pt}{\raggedright 
{\scriptsize 0.49}
} & \parbox{21pt}{\raggedright 
{\scriptsize 8.01}
} & \parbox{21pt}{\raggedright 
{\scriptsize 0.76}
} & \parbox{21pt}{\raggedright 
{\scriptsize 3.83}
} & \parbox{21pt}{\raggedright 
{\scriptsize 0.69}
} \\
\hline
\parbox{51pt}{\raggedright 
{\scriptsize A2142}
} & \parbox{14pt}{\centering 
{\scriptsize 9.70}
} & \parbox{56pt}{\centering 
{\scriptsize 0.27}
} & \parbox{28pt}{\raggedright 
{\scriptsize 0.591}
} & \parbox{28pt}{\raggedright 
{\scriptsize 108.5}
} & \parbox{28pt}{\raggedright 
{\scriptsize 2537}
} & \parbox{35pt}{\raggedright 
{\scriptsize 2.39}
} & \parbox{21pt}{\raggedright 
{\scriptsize 15.93}
} & \parbox{21pt}{\raggedright 
{\scriptsize 2.32}
} & \parbox{21pt}{\raggedright 
{\scriptsize 4.36}
} & \parbox{21pt}{\raggedright 
{\scriptsize 1.37}
} \\
\hline
\parbox{51pt}{\raggedright 
{\scriptsize A2244}
} & \parbox{14pt}{\centering 
{\scriptsize 7.10}
} & \parbox{56pt}{\centering 
{\scriptsize 0.23}
} & \parbox{28pt}{\raggedright 
{\scriptsize 0.607}
} & \parbox{28pt}{\raggedright 
{\scriptsize 88.7}
} & \parbox{28pt}{\raggedright 
{\scriptsize 1773}
} & \parbox{35pt}{\raggedright 
{\scriptsize 0.84}
} & \parbox{21pt}{\raggedright 
{\scriptsize 8.36}
} & \parbox{21pt}{\raggedright 
{\scriptsize 0.92}
} & \parbox{21pt}{\raggedright 
{\scriptsize 2.45}
} & \parbox{21pt}{\raggedright 
{\scriptsize 0.71}
} \\
\hline
\parbox{51pt}{\raggedright 
{\scriptsize UGC03957}
} & \parbox{14pt}{\centering 
{\scriptsize 2.58}
} & \parbox{56pt}{\centering 
{\scriptsize 0.09}
} & \parbox{28pt}{\raggedright 
{\scriptsize 0.740}
} & \parbox{28pt}{\raggedright 
{\scriptsize 100.0}
} & \parbox{28pt}{\raggedright 
{\scriptsize 764}
} & \parbox{35pt}{\raggedright 
{\scriptsize 0.08}
} & \parbox{21pt}{\raggedright 
{\scriptsize 1.57}
} & \parbox{21pt}{\raggedright 
{\scriptsize 0.09}
} & \parbox{21pt}{\raggedright 
{\scriptsize 0.47}
} & \parbox{21pt}{\raggedright 
{\scriptsize 0.13}
} \\
\hline
\parbox{51pt}{\raggedright 
{\scriptsize S636}
} & \parbox{14pt}{\centering 
{\scriptsize 1.18}
} & \parbox{56pt}{\centering 
{\scriptsize 0.01}
} & \parbox{28pt}{\raggedright 
{\scriptsize 0.752}
} & \parbox{28pt}{\raggedright 
{\scriptsize 242.3}
} & \parbox{28pt}{\raggedright 
{\scriptsize 742}
} & \parbox{35pt}{\raggedright 
{\scriptsize 0.06}
} & \parbox{21pt}{\raggedright 
{\scriptsize 0.65}
} & \parbox{21pt}{\raggedright 
{\scriptsize 0.03}
} & \parbox{21pt}{\raggedright 
{\scriptsize 0.09}
} & \parbox{21pt}{\raggedright 
{\scriptsize 0.05}
} \\
\hline
\parbox{51pt}{\raggedright 
{\scriptsize M49}
} & \parbox{14pt}{\centering 
{\scriptsize 0.95}
} & \parbox{56pt}{\centering 
{\scriptsize 0.26}
} & \parbox{28pt}{\raggedright 
{\scriptsize 0.592}
} & \parbox{28pt}{\raggedright 
{\scriptsize 7.7}
} & \parbox{28pt}{\raggedright 
{\scriptsize 177}
} & \parbox{35pt}{\raggedright 
{\scriptsize 0.001}
} & \parbox{21pt}{\raggedright 
{\scriptsize 0.109}
} & \parbox{21pt}{\raggedright 
{\scriptsize 0.002}
} & \parbox{21pt}{\raggedright 
{\scriptsize 0.041}
} & \parbox{21pt}{\raggedright 
{\scriptsize 0.009}
} \\
\hline
\parbox{51pt}{\raggedright 
{\scriptsize A1689}
} & \parbox{14pt}{\centering 
{\scriptsize 9.23}
} & \parbox{56pt}{\centering 
{\scriptsize 0.33}
} & \parbox{28pt}{\raggedright 
{\scriptsize 0.690}
} & \parbox{28pt}{\raggedright 
{\scriptsize 114.8}
} & \parbox{28pt}{\raggedright 
{\scriptsize 1898}
} & \parbox{35pt}{\raggedright 
{\scriptsize 1.23}
} & \parbox{21pt}{\raggedright 
{\scriptsize 13.21}
} & \parbox{21pt}{\raggedright 
{\scriptsize 1.61}
} & \parbox{21pt}{\raggedright 
{\scriptsize 5.14}
} & \parbox{21pt}{\raggedright 
{\scriptsize 1.13}
} \\
\hline
\parbox{51pt}{\raggedright 
{\scriptsize A1800}
} & \parbox{14pt}{\centering 
{\scriptsize 4.02}
} & \parbox{56pt}{\centering 
{\scriptsize 0.04}
} & \parbox{28pt}{\raggedright 
{\scriptsize 0.766}
} & \parbox{28pt}{\raggedright 
{\scriptsize 276.1}
} & \parbox{28pt}{\raggedright 
{\scriptsize 1284}
} & \parbox{35pt}{\raggedright 
{\scriptsize 0.34}
} & \parbox{21pt}{\raggedright 
{\scriptsize 4.14}
} & \parbox{21pt}{\raggedright 
{\scriptsize 0.36}
} & \parbox{21pt}{\raggedright 
{\scriptsize 1.15}
} & \parbox{21pt}{\raggedright 
{\scriptsize 0.38}
} \\
\hline
\end{tabular}
\vspace{2pt}

\vspace{3pt} \noindent
\begin{tabular}{|p{51pt}|p{14pt}|p{56pt}|p{28pt}|p{28pt}|p{28pt}|p{28pt}|p{35pt}|p{35pt}|p{35pt}|p{35pt}|}
\hline
\parbox{51pt}{\raggedright 
{\scriptsize A1914}
} & \parbox{14pt}{\centering 
{\scriptsize 10.5}
} & \parbox{56pt}{\centering 
{\scriptsize 0.22}
} & \parbox{28pt}{\raggedright 
{\scriptsize 0.751}
} & \parbox{28pt}{\raggedright 
{\scriptsize 162.7}
} & \parbox{28pt}{\raggedright 
{\scriptsize 1768}
} & \parbox{35pt}{\raggedright 
{\scriptsize 1.08}
} & \parbox{21pt}{\raggedright 
{\scriptsize 15.21}
} & \parbox{21pt}{\raggedright 
{\scriptsize 1.79}
} & \parbox{21pt}{\raggedright 
{\scriptsize 7.44}
} & \parbox{21pt}{\raggedright 
{\scriptsize 1.31}
} \\
\hline
\parbox{51pt}{\raggedright 
{\scriptsize NGC5813}
} & \parbox{14pt}{\centering 
{\scriptsize 0.52}
} & \parbox{56pt}{\centering 
{\scriptsize 0.18}
} & \parbox{28pt}{\raggedright 
{\scriptsize 0.766}
} & \parbox{28pt}{\raggedright 
{\scriptsize 17.6}
} & \parbox{28pt}{\raggedright 
{\scriptsize 166}
} & \parbox{35pt}{\raggedright 
{\scriptsize 0.001}
} & \parbox{21pt}{\raggedright 
{\scriptsize 0.072}
} & \parbox{21pt}{\raggedright 
{\scriptsize 0.001}
} & \parbox{21pt}{\raggedright 
{\scriptsize 0.021}
} & \parbox{21pt}{\raggedright 
{\scriptsize 0.006}
} \\
\hline
\parbox{51pt}{\raggedright 
{\scriptsize NGC5846}
} & \parbox{14pt}{\centering 
{\scriptsize 0.82}
} & \parbox{56pt}{\centering 
{\scriptsize 0.47}
} & \parbox{28pt}{\raggedright 
{\scriptsize 0.599}
} & \parbox{28pt}{\raggedright 
{\scriptsize 4.9}
} & \parbox{28pt}{\raggedright 
{\scriptsize 152}
} & \parbox{35pt}{\raggedright 
{\scriptsize 0.001}
} & \parbox{21pt}{\raggedright 
{\scriptsize 0.082}
} & \parbox{21pt}{\raggedright 
{\scriptsize 0.001}
} & \parbox{21pt}{\raggedright 
{\scriptsize 0.031}
} & \parbox{21pt}{\raggedright 
{\scriptsize 0.007}
} \\
\hline
\parbox{51pt}{\raggedright 
{\scriptsize A2151w}
} & \parbox{14pt}{\centering 
{\scriptsize 2.40}
} & \parbox{56pt}{\centering 
{\scriptsize 0.16}
} & \parbox{28pt}{\raggedright 
{\scriptsize 0.564}
} & \parbox{28pt}{\raggedright 
{\scriptsize 47.9}
} & \parbox{28pt}{\raggedright 
{\scriptsize 957}
} & \parbox{35pt}{\raggedright 
{\scriptsize 0.12}
} & \parbox{21pt}{\raggedright 
{\scriptsize 1.42}
} & \parbox{21pt}{\raggedright 
{\scriptsize 0.09}
} & \parbox{21pt}{\raggedright 
{\scriptsize 0.25}
} & \parbox{21pt}{\raggedright 
{\scriptsize 0.12}
} \\
\hline
\parbox{51pt}{\raggedright 
{\tiny TRIANGULUM}
} & \parbox{14pt}{\centering 
{\scriptsize 9.60}
} & \parbox{56pt}{\centering 
{\scriptsize 0.1}
} & \parbox{28pt}{\raggedright 
{\scriptsize 0.61}
} & \parbox{28pt}{\raggedright 
{\scriptsize 196.5}
} & \parbox{28pt}{\raggedright 
{\scriptsize 2385}
} & \parbox{35pt}{\raggedright 
{\scriptsize 1.98}
} & \parbox{21pt}{\raggedright 
{\scriptsize 15.22}
} & \parbox{21pt}{\raggedright 
{\scriptsize 2.11}
} & \parbox{21pt}{\raggedright 
{\scriptsize 4.48}
} & \parbox{21pt}{\raggedright 
{\scriptsize 1.32}
} \\
\hline
\parbox{51pt}{\raggedright 
{\tiny OPHIUCHUS}
} & \parbox{14pt}{\centering 
{\scriptsize 10.3}
} & \parbox{56pt}{\centering 
{\scriptsize 0.13}
} & \parbox{28pt}{\raggedright 
{\scriptsize 0.747}
} & \parbox{28pt}{\raggedright 
{\scriptsize 196.5}
} & \parbox{28pt}{\raggedright 
{\scriptsize 1701}
} & \parbox{35pt}{\raggedright 
{\scriptsize 0.91}
} & \parbox{21pt}{\raggedright 
{\scriptsize 14.11}
} & \parbox{21pt}{\raggedright 
{\scriptsize 1.59}
} & \parbox{21pt}{\raggedright 
{\scriptsize 6.91}
} & \parbox{21pt}{\raggedright 
{\scriptsize 1.21}
} \\
\hline
\parbox{51pt}{\raggedright 
{\scriptsize ZwC174}
} & \parbox{14pt}{\centering 
{\scriptsize 5.23}
} & \parbox{56pt}{\centering 
{\scriptsize 0.1}
} & \parbox{28pt}{\raggedright 
{\scriptsize 0.717}
} & \parbox{28pt}{\raggedright 
{\scriptsize 163.4}
} & \parbox{28pt}{\raggedright 
{\scriptsize 1354}
} & \parbox{35pt}{\raggedright 
{\scriptsize 0.43}
} & \parbox{21pt}{\raggedright 
{\scriptsize 5.49}
} & \parbox{21pt}{\raggedright 
{\scriptsize 0.50}
} & \parbox{21pt}{\raggedright 
{\scriptsize 1.79}
} & \parbox{21pt}{\raggedright 
{\scriptsize 0.47}
} \\
\hline
\parbox{51pt}{\raggedright 
{\scriptsize A3888}
} & \parbox{14pt}{\centering 
{\scriptsize 8.84}
} & \parbox{56pt}{\centering 
{\scriptsize 0.1}
} & \parbox{28pt}{\raggedright 
{\scriptsize 0.928}
} & \parbox{28pt}{\raggedright 
{\scriptsize 282.4}
} & \parbox{28pt}{\raggedright 
{\scriptsize 1455}
} & \parbox{35pt}{\raggedright 
{\scriptsize 0.71}
} & \parbox{21pt}{\raggedright 
{\scriptsize 12.61}
} & \parbox{21pt}{\raggedright 
{\scriptsize 1.33}
} & \parbox{21pt}{\raggedright 
{\scriptsize 7.14}
} & \parbox{21pt}{\raggedright 
{\scriptsize 1.08}
} \\
\hline
\parbox{51pt}{\raggedright 
{\scriptsize HGC94}
} & \parbox{14pt}{\centering 
{\scriptsize 3.45}
} & \parbox{56pt}{\centering 
{\scriptsize 0.11}
} & \parbox{28pt}{\raggedright 
{\scriptsize 0.514}
} & \parbox{28pt}{\raggedright 
{\scriptsize 60.6}
} & \parbox{28pt}{\raggedright 
{\scriptsize 1237}
} & \parbox{35pt}{\raggedright 
{\scriptsize 0.24}
} & \parbox{21pt}{\raggedright 
{\scriptsize 2.40}
} & \parbox{21pt}{\raggedright 
{\scriptsize 0.19}
} & \parbox{21pt}{\raggedright 
{\scriptsize 0.43}
} & \parbox{21pt}{\raggedright 
{\scriptsize 0.21}
} \\
\hline
\parbox{51pt}{\raggedright 
{\scriptsize RXJ2344}
} & \parbox{14pt}{\centering 
{\scriptsize 4.73}
} & \parbox{56pt}{\centering 
{\scriptsize 0.07}
} & \parbox{28pt}{\raggedright 
{\scriptsize 0.807}
} & \parbox{28pt}{\raggedright 
{\scriptsize 212.0}
} & \parbox{28pt}{\raggedright 
{\scriptsize 1222}
} & \parbox{35pt}{\raggedright 
{\scriptsize 0.34}
} & \parbox{21pt}{\raggedright 
{\scriptsize 4.97}
} & \parbox{21pt}{\raggedright 
{\scriptsize 0.43}
} & \parbox{21pt}{\raggedright 
{\scriptsize 1.78}
} & \parbox{21pt}{\raggedright 
{\scriptsize 0.43}
} \\
\hline
\end{tabular}
\vspace{2pt}

By comparing columns 9 and 11 in Table 1, MOG
and $\kappa{}$-model provide  reasonably close values to each other for the masses 
integrated to rout, respectively  $M_{MSTG}$ and $M_{\kappa{}}$. By comparing columns 7 and 11 in Table 1, the agreement between  $M_{\kappa{}}$
and  $M_{gas}$ is fairly good, and it is clear that,  in most
cases,  the $\kappa{}$-model  does not necessitate dark matter in the outer
regions of the galaxy clusters. In Table 1, $M_{\kappa{}}$ is smaller than $M_{MSTG}$  when the  temperatures are  higher than $5-6\ keV$,  whereas  the reverse is true  when the temperatures are lower than $5-6\ keV$.   On the other hand, for  $T \sim 5-6\ keV$, then $M_{\kappa{}} \sim M_{MSTG}$. Eventually, when the temperatures are smaller than $1 \ keV$, then    $M_{\kappa{}}$ are   larger than $M_{MSTG}$.

By contrast,  convergent MOND  dynamic mass, $M_{MOND}$,   are systematically  located too high.  Even though MOND substantially decreases the dark matter content for convergent
MOND dynamic mass, this theory is not able   to completely  remove  dark
matter in the galaxy clusters. By comparing the columns 10 ($M_{MOND}$) and 7 ($M_{gas}$) of Table
1, $M_{MOND}$ is sometimes superior by
 a mean  factor of the order of  $2$ to  $M_{gas}$.   Galaxy cluster
stability  cannot be fully explained by the current MOND formulation alone. MOND, at least
its initial form  (Milgrom, 1983), still  needs to use a residual  content of invisible  matter  for galaxy clusters.
However, owing to the great  success of MOND for the individual galaxies, a proper
direction is to build  a multiscale  MOND, which is able to  simultaneously explain both  individual
galaxies and   galaxy clusters.  Fortunately,  this model has been 
built.  This is the main motivation of EMOND;   EMOND assumes that there is an  increase in the fundamental parameter $a_0$ of the MOND paradigm  in galaxy clusters   compared to the value selected  for this coefficient  in the case of  individual galaxies  (Zhao and Famaey, 2012; Hodson and Zhao, 2017 a, b).  This rescaling adapts   the parameter  $a_0$ to  the size  of the object under consideration and appears to be quite  natural. In this case,  MOND can eventually and adequately  fit the ICM  gas mass integrated to $r_{out}$. Note, in the context of the $\kappa$-model  a physical interpretation of this rescaling in MOND is supplied. The mean density in   a galaxy cluster is much smaller than the corresponding one  in an individual  galaxy by a factor $\sim 40$ (center of the cluster) and $\sim 200-1000$ (in the outskirts of the cluster). In our study, the rescaling is given by an estimate of  $\kappa$,   and $\kappa_G$/$\kappa_{cluster} \sim 5-10$.

\vspace{5pt}

Despite these promising considerations for the $\kappa$-model,   MOG  appears to be slightly better as  for the fitting of the observational curves. We might argue  that MOG has two free outer parameters, whereas the  $\kappa$-model has none; therefore, it is easier to fit a curve with a sufficient number of outer free parameters than without any free parameter.  However, on closer  examination of MOG, we can see that a multifit procedure   is used  throughout the calculations. MOG  relies on two parameters $M_0$ and  $r_0$, but these parameters are not constant across the series of galaxy clusters. Moreover, $M_0$ depends on the ICM gas mass for each individual clusters (Brownstein and Moffat, 2006, paragraph 4). Thus it seems that in MOG the results are already  included at the start of the calculations, and the method appears  somewhat post hoc. Thus, MOG is not a fully ab initio procedure. Due to these reasons,   the $\kappa$-model and  EMOND are more efficient than MOG as for their predictive  power.

\section{Bullet Cluster}

The Bullet Cluster (1E 0657-56)  is very often presented as a clear proof of the
existence of dark matter (Figure 3). The Bullet Cluster is composed of two
colliding clusters: a main cluster (Mc) and a small or sub-cluster (Sc) (the bullet per se)
(Brada\v{c} el al, 2005, 2006).

\begin{center}
\includegraphics[height=200pt, width=210pt]{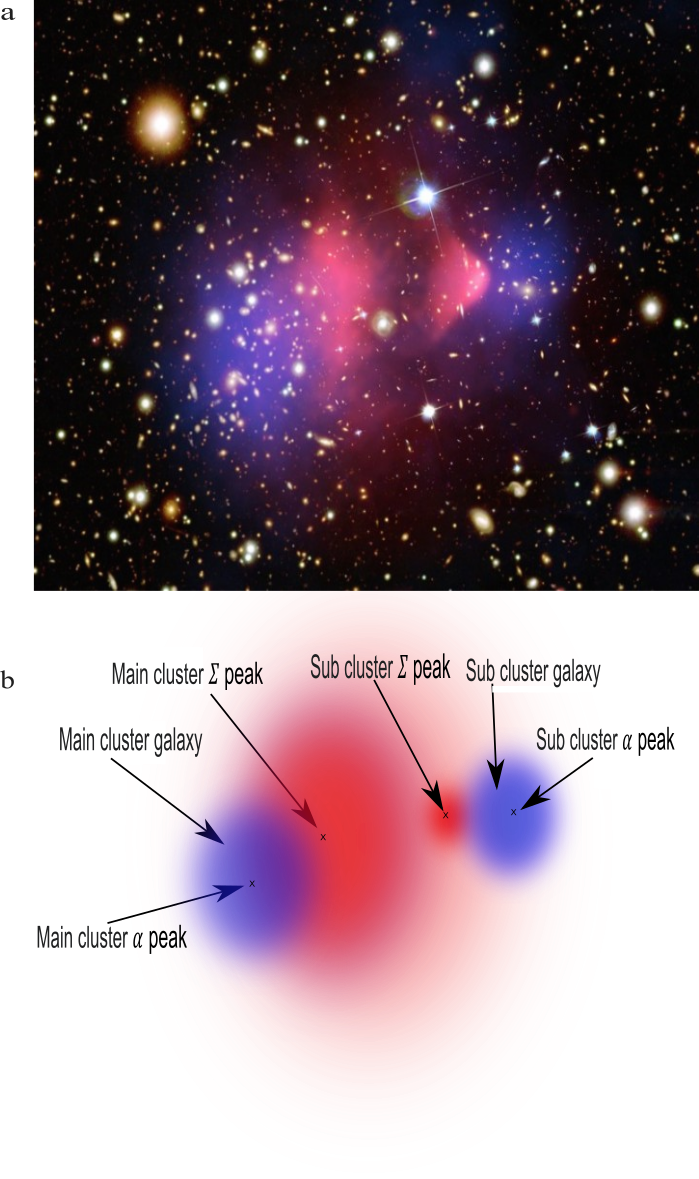}

Figure 4
\end{center}

To determine how the $\kappa$-model reinterprets the observational data for the
Bullet Cluster, we initially start with the known (but apparent) surface densities  for both  the
hot gas and the visible galaxies (Figure 5). Figures 5a and 5b are reproduced from 
Brownstein and Moffat (2006). In addition, Table 2 provides the  masses for the different 
components. The assumed  DM  content is adjusted  such that  the total mass ratio  $DM/B$ is $ \sim 6$.

\begin{center}
\includegraphics[height=100pt, width=310pt]{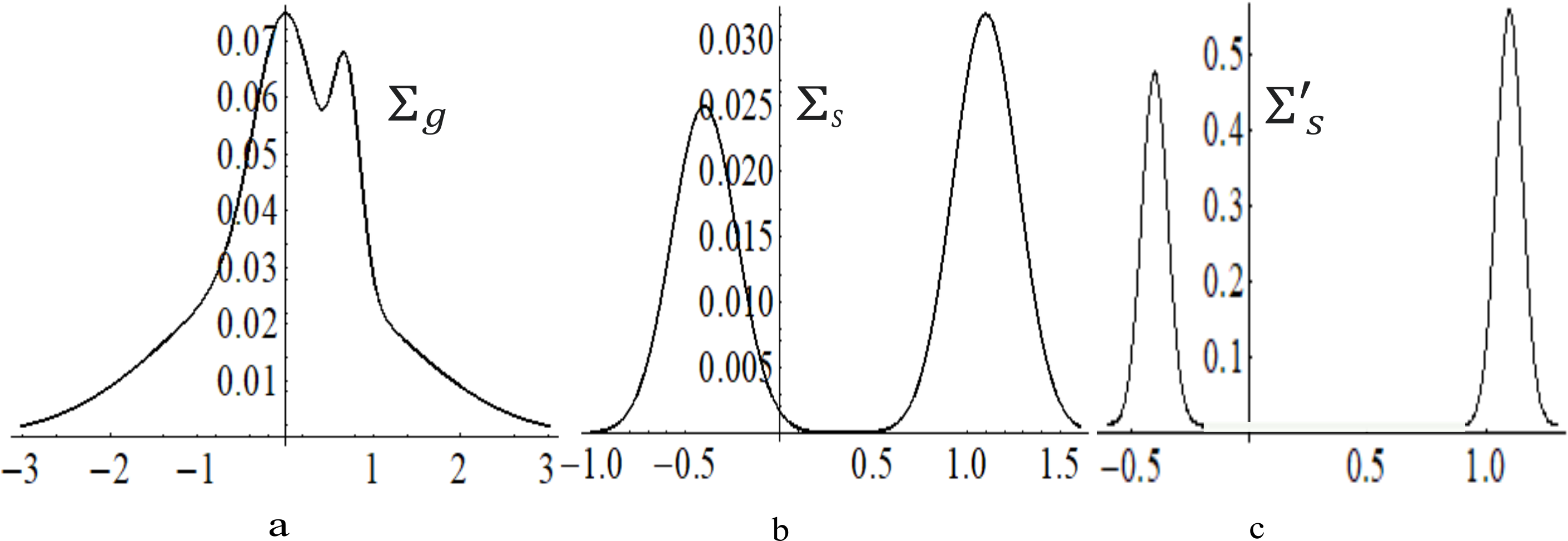}

Figure 5
\end{center}

\noindent 
Figure 5 Abscissa : distances, unit $500\  kpc$.  Ordinate:
surface densities,  unit   $3.1\ {10}^3\ M_\odot\ {pc}^{-2}$.
The abscissa axis  passes through the\ galaxy cluster centers.   Figure 5a: scaled plot of hot gas
density, Figure 5b: scaled plot of apparent galaxy density observed from Earth (Main cluster + Bullet) and Figure 5c: scaled plot of real galaxy
density measured by a hypothetical  observer located inside the  Main or the Sub group of
visible galaxies (represented by blue disks in Figure 4). The mass is 
invariable by  the transform: apparent $\longleftrightarrow$ real surface densities, i.e. $\int dxdy\ {\Sigma{}}_s (x,y)=\int dxdy\ {\Sigma{}}_s' (x,y)$.

{\raggedright

\vspace{3pt} \noindent
\begin{tabular}{|p{63pt}|p{96pt}|p{80pt}|p{80pt}|p{79pt}|}
\hline
\parbox{63pt}{\centering 
{\footnotesize Component}
} & \parbox{96pt}{\centering 
{\footnotesize Main cluster 
  (Main)}
} & \parbox{80pt}{\centering 
{\footnotesize Subcluster    (Sub)}
} & \parbox{80pt}{\centering  
{\footnotesize Diffuse component}
} & \parbox{79pt}{\centering 
{\footnotesize Total}
} \\
\hline
\parbox{63pt}{\centering  ${M}_{gas}$} & \parbox{96pt}{\centering 
$4.38\ {10}^{13}$
} & \parbox{80pt}{\centering  $1.93\ {10}^{13}$} & \parbox{80pt}{\centering  
$8.01\ {10}^{13}$
} & \parbox{79pt}{\centering  $1.43\ {10}^{14}$
} \\
\hline
\parbox{63pt}{\centering 

$M_{galaxies}$

} & \parbox{96pt}{\centering 
$4.67\ {10}^{12}$
} & \parbox{80pt}{\centering 
$3.46\  {10}^{12}$
} & \parbox{80pt}{\centering  --} & \parbox{79pt}{\centering

$8.32\  {10}^{12}$
} \\
\hline
\parbox{63pt}{\centering  $M_{DM}$} & \parbox{96pt}{\centering 
$6.54\ {10}^{14}$
} & \parbox{80pt}{\centering 
$1.67\ {10}^{14}$

} & \parbox{80pt}{\centering --} & \parbox{79pt}{\centering

$8.21\ {10}^{14}$
} \\
\hline
\end{tabular}
\vspace{2pt}
}

\begin{center}
Table 2 The masses are expressed in  solar mass $M_\odot$
\end{center}

 In the usual treatment of the galaxy clusters (paragraph 2), only the gaseous
component are considered (the stellar fraction is negligible in 
galaxy clusters). In the specific case of the Bullet Cluster, the situation is very different. We have two
clearly identified parts:  a very massive gaseous component and  a galactic 
component; however, these two components are separated. Based on the lensing diagram, only the gaseous component needs to be considered in the calculation of $\kappa{}$. Thus, if we assume that the Main and
Sub  groups of visible galaxies  (the disks displayed in blue in
Figure 6)  have a low content of gas,  the  lensing due to the $\kappa$ effect  is much
higher than that in the hot gas (the disk displayed  in red in
Figure 6). The usual relationship of the  $\kappa{}$-model is as follows:

\begin{equation}
\frac{{\kappa{}}_M}{\kappa{}}=1+Ln\
\left(\frac{{\Sigma{}}_M}{{\Sigma{}}_{gas}}\right)
\end{equation}

\noindent where ${\Sigma{}}_M \sim{} 0.075\times{}3.1\ {10}^3\ M_\odot\ {pc}^{-2}=232.5 \ M_\odot\
{pc}^{-2}$   is the maximal value taken   by the surface density  of hot
gas (Figure 5a).  Figure 5a can be used to provide the following relationship:

\begin{equation}
{({\Sigma{}}_{gas}^{Main})}_{outer} \ , \  {({\Sigma{}}_{gas}^{Sub})}_{outer}\
\sim{} 0.7\times{} 232.5\ M_\odot\ {pc}^{-2}=162.7\ M_\odot\ {pc}^{-2}
\end{equation}

\noindent The latter   quantity represents the  surface density of the hot gas located in the immediate
environment of  the Main and Sub groups of visible galaxies; each of them being displayed by a blue
disk  in Figure 4.

 However, following the  current interpretation, both the Main and Sub galaxy clusters were stripped  from the hot gas  during  the 
collision. The amount of the gas remaining  inside
the groups of visible galaxies after this  collision needs to be determined. Clearly the hot gas initially located inside the Main and Sub groups of visible galaxies is removed by the strong
shock during the collision, and each of these groups  is now located  in a subdense bubble with a low content of hot gas. A
pre-analysis of the $\alpha{}$-diagram in the $\kappa$-model context, compared to that
provided by observations (Brada\v{c} el al, 2005, 2006), enables to predict that  
the amount of hot gas inside  the subdense  bubbles is  $\sim 15\%$ of the amount of hot gas surrounding them,  i.e. ${({\Sigma{}}_{gas}^{Main})}_{inner}$ or
${({\Sigma{}}_{gas}^{Sub})}_{inner}=24.4\ M_\odot\ {pc}^{-2}$. With 
equal temperatures, the gas pressure in
the groups of visible galaxies  (blue regions in the Figure 4) is thus predicted to be lower 
than that in the immediate  region surrounding them (red region).  Evidently, the
system  is not in hydrostatic equilibrium, and  this situation
cannot indefinitely persist.   The required time  to re-equalize the densities
is approximately $\frac{R_c}{3c_s}\sim{}100\
Myears$, where  $R_c\sim{}0.5\ Mpc$ is the mean characteristic size of the Main or Sub  groups of visible galaxies  and $c_s$ is the speed of sound in  the hot gas ($T_e\sim{}{10}^8\ K)$\footnote{After this short  period of time, the $\kappa$-model predicts that the lensing diagram  will no longer be centered on the groups  of visible galaxies, but on the hot gas and  will likely resemble Figure 7a.  Thus, we see an instantaneous phase of a rapidly evolutionary process.}.   With the relationship
(10) and the aforementioned values of  ${({\Sigma{}}_{gas}^{Main})}_{inner}$ or
${({\Sigma{}}_{gas}^{Sub})}_{inner}$, we can determine the
amplification factor $\kappa{}$ inside  the two groups of visible galaxies as follows:

\begin{equation}
\frac{{\kappa{}}_M}{{\kappa{}}_{gas}^{Main}}=\frac{{\kappa{}}_M}{{\kappa{}}_{gas}^{Sub}}\sim{}\frac{{\kappa{}}_M}{{\kappa{}}_{gas}^{galactic}}=3
\end{equation}

\begin{center}
\includegraphics[height=210pt, width=450pt]{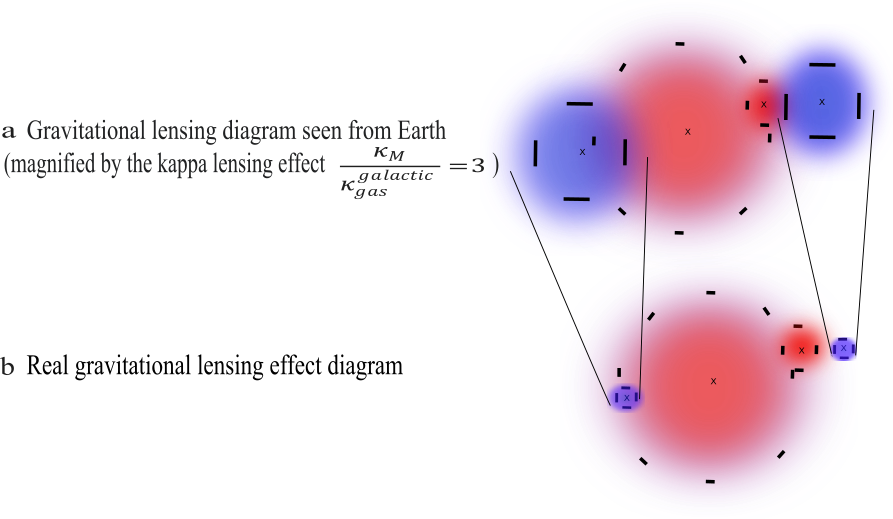}
\end{center}

\noindent Figure 6  Basic illustration of the combined action of both  gravitational and kappa lensings. The
lensing by the hot gas is used  as a reference for the two figures a and b.

\vspace{10pt}
 In Figure 6a,  the combined product of both the gravitational and
kappa lensing effects is clearly centered on the Main and Sub groups of visible galaxies. In Figure 6b, as expected, the lensing is clearly
centered on the hot gas (which contains $90\%$ of  the baryonic mass against $10\%$
for the visible galaxies in a galaxy cluster).  However, the reality shown in
 Figure 6b  cannot be viewed by a unique observer (for instance, a terrestrial
observer surrounded by its own environment). Specifically, two distinct  observers are needed  to determine the effect. 
The  part located in the blue area (groups  of visible galaxies) is perceived   by a
hypothetical  observer situated inside any region where the mean gas density is the same as  that in  the Main or Sub group  of visible galaxies;  the part
located in the red area (hot gas)  is perceived by a hypothetical  observer 
situated inside any region where the mean density is the same as  that in 
the hot gas. If 
 the $\kappa{}$-model paradigm is expected to be on the right track, then the Bullet Cluster  is an illustrative (but rare) example that the perception of the objects in the Universe is observer-dependent.

Beyond these qualitative considerations, calculations are evidently  needed to ascertain our purpose. The procedure is
well known.  The field equations of general relativity can be linearized if the
gravitational field is weak. Then, the deflection angle of a set of masses is
simply the vectorial sum of the deflections due to individual lenses. The
plane of the sky is $(x,\ y)$. The deviation angle $\boldsymbol{\alpha{}}$ can be written
using the thin lens approximation as follows(Bartelmann and Schneider, 2001):

\begin{equation}
\boldsymbol{\alpha{}}\left(x,y\right)=\int_Sdx’dy’\frac{4G}{c}\Sigma{}(x’,y’)\frac{\bold{r}-\bold{r’}}{{\left\vert{}\bold{r}-\bold{r}’\right\vert{}}^2}
\end{equation}

where $G$ is the gravitational constant and $c$ is the speed of light. The density distribution $\Sigma{}$ is
integrated  over  all the surface $S$ of the cluster system. The  total
surface density $\Sigma{}\left(x,y\right)$ is the sum of the gas  and stellar
surface densities.

\begin{equation}
\Sigma{}\left(x,y\right)={\Sigma{}}_g\left(x,y\right)+{\Sigma{}}_s’\left(x,y\right)
\end{equation}

 where ${\Sigma{}}_g\left(x,y\right)$  and ${\Sigma{}’}_s\left(x,y\right)$ are the fits of the
distributions represented in Figures 4a and 4c, respectively; these are assumed to be
approximately circular and Gaussian.

\noindent We know that the $\kappa{}$-model does not change the local physics, apart from a
magnification factor locally applied. Thus, in the  $\kappa{}$-model, the
relationship (14) eventually  becomes the following:\footnote{A multiplicative  factor,
$\frac{{\kappa{}}_E}{{\kappa{}}_M}$,  needs to be applied to
$\bold{\alpha{}}\left(x,y\right)$; however, the factor is a global and constant coefficient
independent of $x,y$, which does not affect the relative position of the peaks on the
lensing diagram.} 

\begin{equation}
\boldsymbol{\alpha{}}\left(x,y\right)=\int_Sdx’dy’\frac{4G}{c}\left[{\Sigma{}}_g\left(x,y\right)+\frac{{\kappa{}}_M}{{\kappa{}}_{gas}^{galactic}}{\Sigma{}’}_s\left(x,y\right)\right]\frac{\bold{r}-\bold{r}’}{{\left\vert{}\bold{r}-\bold{r}’\right\vert{}}^2}
\end{equation}

 Figure 7 shows the lensing diagrams  $\left\vert{}\boldsymbol{\alpha{}}\right\vert{}$ of both dark matter
and $\kappa{}$-model and Figure 8 
shows the components ${\alpha{}}_x$ and ${\alpha{}}_y$. Even though
the physical  interpretation is very different,  the $\kappa{}$-model
and  DM diagrams are very similar and both show that the lensing is centered
on the visible galaxies and not on the hot gas,  which is centered at $(0,0)$. The
essential difference  between the two models is  that the $\kappa{}$-model does not
need dark matter to explain the same facts.

\begin{center}
\includegraphics[height=460pt, width=470pt]{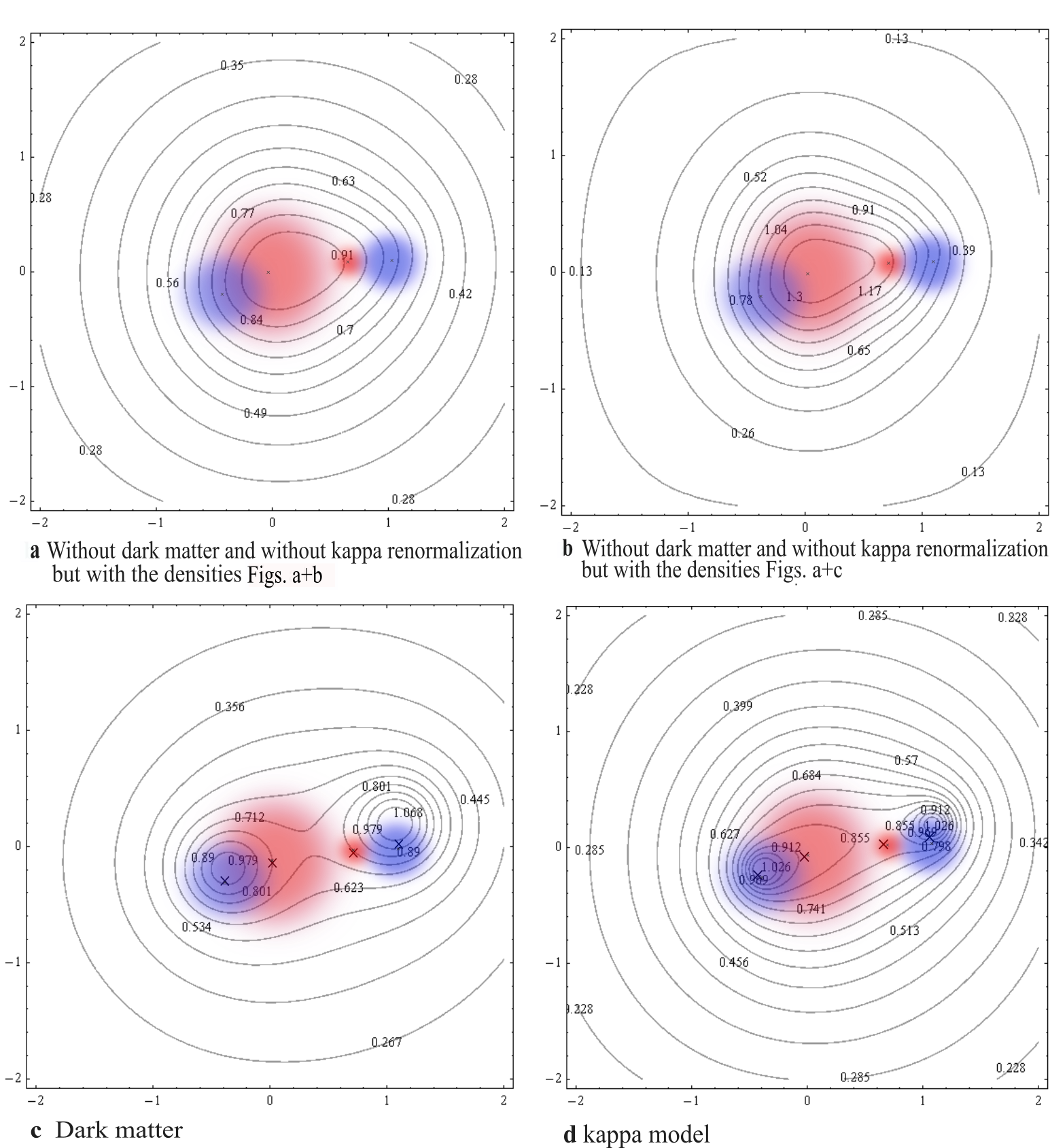}
\end{center}

\noindent 
Figure 7 Comparative plots of the lensing diagram. The crosses indicate the
positions of the centers of the different distributions (hot gas in red and groups of visible galaxies in blue)

\begin{center}
\includegraphics[height=460pt, width=480pt]{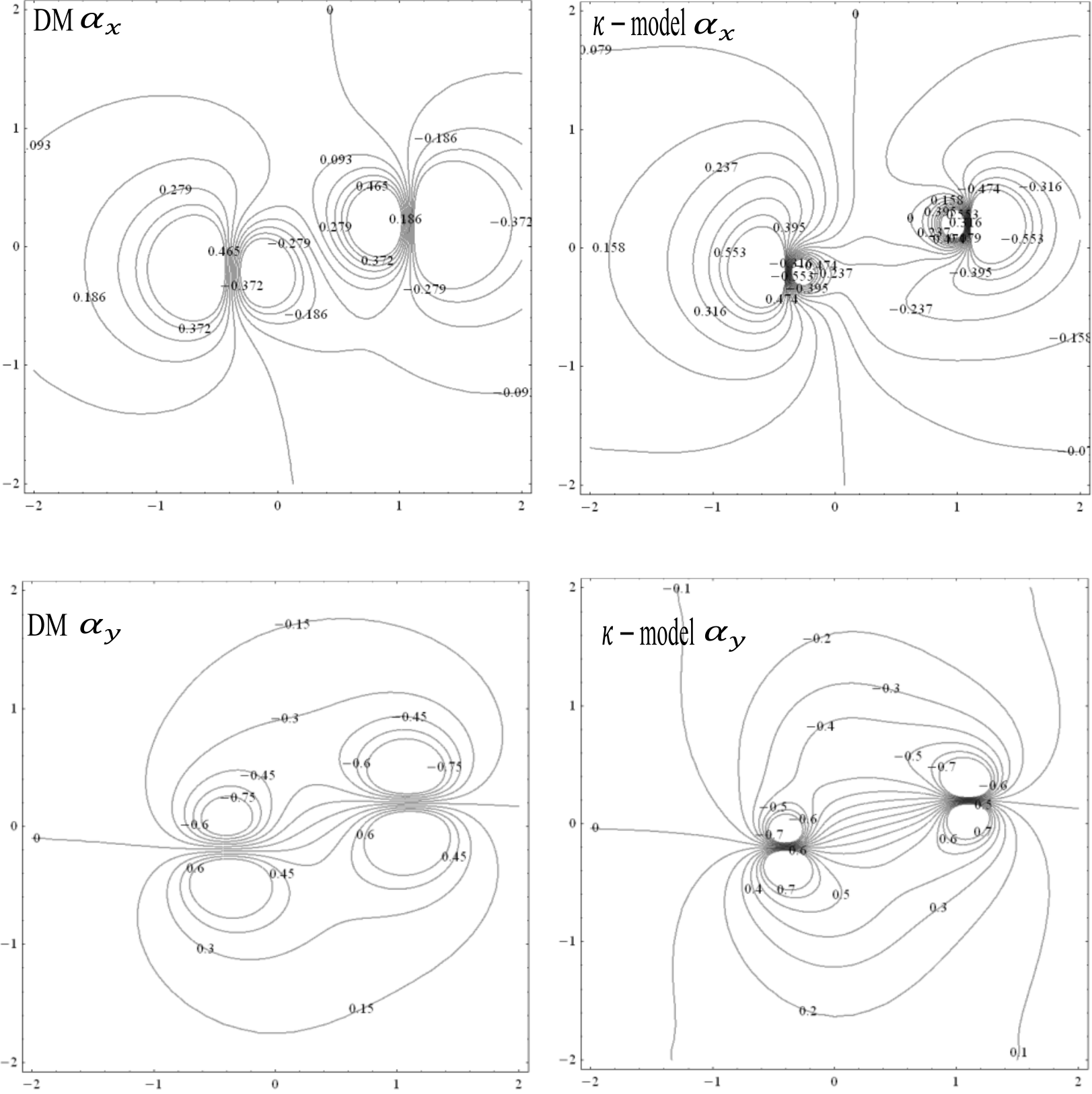}

Figure 8  The same as Figure 7 but for the components of $\boldsymbol{\alpha}$
\end{center}

The method applied to the Bullet Cluster can be extended to other similar cases such as the Train Wreck Cluster (Abell 620). In the latter situation the lensing is centered on the hot gas. Even though the morphology of the Train Wreck Cluster is much more complex than that of the Bullet Cluster (Jee et al, 2014), a natural explanation in the framework of the $\kappa$-model is that  the intergalactic gas filling rate  of the  bubbles containing the visible galaxies is different. With a filling rate of $60\%$ instead of $15\%$ (Bullet Cluster)  we have been able to ascertain that the lensing is no longer centered on the visible galaxies.

\section{Conclusion}

Effectively tested MOND and MOG models were designed to understand the
dynamics of the Universe without dark matter. By contrast, the dark
matter paradigm is an ad hoc concept, where the dark matter content has to be
adapted to each situation and consequently, has no predictive value. However, and paradoxically enough, the DM paradigm constitutes a near unanimous recognition among  the astrophysicists. The main reason is that the DM  paradigm is very easy to use and  effectively works at all times due to its extreme flexibility.

On the other hand, in  the framework of the  $\kappa{}$-model, the single issue  of
the baryonic mass needs to be sufficient to understand the dynamics of galaxies and that of
the galaxy clusters, without dark matter or artificial  ingredients, such as the
introduction of new  parameters into the calculations. Specifically, the  $\kappa{}$-model
aims   to determine a one-to-one  relationship directly
 linking the sole observational  data, i.e. the estimated mean density,  to either
 the spectroscopic velocities  in the galaxies or X-ray temperatures in galaxy
clusters.  The $\kappa{}$-model is a MOND-type model, and a behavior very similar
to MOND needs to be found for the galaxy clusters.  In reality, even  though the shape of curves are effectively similar, the $\kappa{}$-model  
curves are strongly parallel-displaced and systematically move 
closer to the observational curves than MOND. The agreement between the $\kappa{}$-model
prediction and the observational data for the total mass of the hot gas in a galaxy cluster is  satisfactory when the  mass  
ratio DM/B  is less than 10, but for values exceeding this
ratio the observational total mass of gas  cannot be effectively predicted  in the framework of an isothermal model.
In this case,  a lower temperature for the hot gas is predicted in the inner regions of the
galaxy clusters.

Finally, the ordered series is as follows: the $\kappa{}$-model has no outer
parameter  (except the internal parameters relative to the system type
(individual galaxies or galaxy clusters), i.e. mean density and temperature); MOND is similar with  just  a single outer parameter; MOG has  two outer parameters (unfortunately, it is in fact a disguised  gas-mass dependent
multifit parameter for the galaxy clusters, i.e. the results are already  included in the hypotheses),  and DM is  an artificial and ad hoc
procedure that  works  at all times. Our proposed logical program was developed to fit the observational curves with only the observational
parameters in a self-consistent manner  (i.e. through  the triplet mean
densities, spectroscopic velocities, and temperatures), and this ultimate goal  is now possible in the
framework of the $\kappa{}$-model.

\vspace{10pt}
{\raggedright{\textbf{Acknowledgment} : the author wishes to thank the anonymous reviewers for useful comments and suggestions.}}

\vspace{10pt}
{\raggedright{\textbf{Data availability statement}: The author confirms that the data supporting the findings of this study are available within the article and the reference list.}}

\vspace{10pt}
{\raggedright{
{\textbf{Conflicts of Interest:} The author declares no conflict of interest.}}}

\section{References}

Banik, I. \& Zhao, H.,  2022, Symmetry, 14, 1331

\noindent 
Bartelmann, M., \& Schneider, P., 2001, Physics Reports, 340, 291

\noindent 
Brada\v{c}, M., Schneider, P., Lombardi, M., \& Erben, T., 2005, A \& A,
436, 39

\noindent 
Brada\v{c}, M., Clowe, D., Gonzalez, A.H., Marshall, P., Forman, W., Jones, C., Markevitch, M., 2006, ApJ, 652, 937

\noindent 
Randall, S., Schrabback, T., \& Zaritsky, D., 2006, ApJ, 652, 937

\noindent 
Browstein, J.R., \& Moffat, J.,  2006, MNRAS, 367, 527

\noindent 
Browstein, J.R., \& Moffat, J., 2007, MNRAS, 382, 29

\noindent 
Bykov, A.M.,   Churazov, E.M.,  Ferrari, C., 
Forman, W.R.  Kaastra, J.S.,  Klein, U.,  
Markevitch, M., \&  J. de Plaa, J., 2015, Space Sci Rev, 188:141

\noindent 
Calcagni, G., Adv. Theor. Math.  Phys., 2012, 16, 549

\noindent 
Capozziello, S., \& De Laurentis, M., 2012, Ann. D. Physik, 524, 545

\noindent 
Cavaliere, A. L. \& Fusco-Femiano, R. 1976, A\&A, 49, 137

\noindent 
DAMA/LIBRA Collaboration, 2022, arXiv.2208.05158

\noindent 
Famaey, B, \& McGaugh, S., 2012, Living Rev. Relativity 15, 10

\noindent 
Freundlich, J., Famaey, B., Oria, P.A.,  B\'{\i}lek, M., M\"{u}ller~O, \& Ibata, R., 2022, A\&A, 658, A24

\noindent 
Giunti, C, \&  Lasserre, T., 2019, Annual Review of Nuclear and Particle Science, Volume 69, 163

\noindent 
Hodson,  A.O,. \& Zhao H., 2017a, A\&A, 598, A127

\noindent 
Hodson,  A.O,. \& Zhao H., 2017b, A\&A, 608, A109

\noindent
Jee, M. J., Hoekstra, H., Mahdavi, A., \& Babul, A., 2014, ApJ. 783 (2): 78

\noindent 
Mc Gaugh, S., 2015, Canadian Journal of Physics. 93~(2): 250

\noindent
McGaugh, S.S., Lelli, F., \& Schombert, J. M. 2016, Physical Review Letters,
117, 201101

\noindent 
Mannheim, P.D., \& Kazanas, D., 1989, ApJ, 342, 635

\noindent 
Milgrom, M. 1983,  Astrophys. J. 270, 365

\noindent 
Milgrom, M., Canadian Journal of Physics,  2015, 93(2): 107

\noindent 
Moffat, J. W.,2006, Journal of Cosmology and Astroparticle Physics, 3, 4

\noindent 
Moffat, J.W.,  2020, Eur. Phys. J. C, 80(10), 906

\noindent 
Niikura, H.,  Takada,  M., Yasuda, N., Lupton, R.H.,  Sumi, T.,  More, S.,
Kurita, T.,  Sugiyama, S., More,  A., Oguri, M., \& Chiba, M., 2019, Nature
Astronomy, 3, 524

\noindent 
O'Brien, J.G., \& Moss, R.J., 2015, J. Phys.: Conf. Ser. 615, 012002

\noindent 
Pascoli, G. \& Pernas, L., 2020, hal: 02530737

\noindent 
Pascoli, G, 2022a, Astrophys. and  Space Sci., 367, 121

\noindent 
Pascoli, G.,  2022b, arXiv: 2205.03062

\noindent 
Reiprich, T.H, 2001, Ph.D.Dissertation, Cosmological Implications and Physical Properties of an X-Ray Flux-Limited Sample of Galaxy
Clusters, Ludwig-Maximilians-Universit\"{a}t M\"{u}nchen

\noindent 
Reiprich, T.H. \& B\"{o}hringer, H. 2002, ApJ, 567, 716

\noindent 
Sch\"{o}del,R. Gallego-Cano, E., Dong,  Nogueras-Lara, F., Gallego-Calvente, A.T., Amaro-Seoane, P. \&  Baumgardt, H., 2018, A\&A, 609, A27

\noindent 
Skordis, C, \& Z\l{}o\'{s}nik, T., 2021,  Phys. Rev. Lett., 127, 161302

\noindent
Varieschi, G.U., 2020, Found Phys 50, 1608

\noindent 
Varieschi, G.U. \&  Calcagni, G., 2022,   JHEP, 2022,  24

\noindent
Varieschi, G.U., 2023, Universe, 9, 246

\noindent 
Verlinde, E.P., 2017, SciPost Phys. 2, 016

\noindent 
Walker, S.,  Simionescu, A.,  Nagai D., Okabe, N.,  Eckert, D.,  Mroczkowski, T., 
Akamatsu, H.,  Ettori, S., \&  Ghirardini, V.,  2019, Space Science Reviews,  215: 7

\noindent 
Watanabe, M., Yamashita, K., Furuzawa, A., Kunieda, H., \& Tawara, Y., 1999,   ApJ, 527: 80

\noindent 
Xenon Collaboration, 2023,  arxiv2303.14729v1

\noindent 
Zhao, H., \& Famaey, B., 2012, Phys. Rev. D, 86, 067301

\vspace{40pt}

\begin{center}
\large{\textbf{Appendix A : Case of an extended set of points}}
\end{center}

As discussed in section 2,  in  the $\kappa{}$-model,  the dynamic
equation, formally written for an infinitesimal element of matter 
of mass $dm$ and subjected to a force $d\boldsymbol{f}$, is as follows:

\begin{equation}
dm\frac{d}{dt}\left(\kappa{}\dot{\boldsymbol{\sigma{}}}\right)=d\boldsymbol{f}
\end{equation}

where $\kappa{}$  is a local  normalization coefficient applied to the spatial  lengths and $d\boldsymbol{f}$  is  measured in situ, i.e. where the element of matter resides. 
However, this equation has no meaning without a reference frame
predefined in advance. First, the internal motions of any  particle (a star, a
nebula) in a galaxy needs to be  studied in the reference frame $R_A$, in which the
barycenter of the galaxy is at rest. Such a reference frame can be 
built by considering a collection $S_{\left\{A_i\right\}}$  of open sets $U_{A_i}$ which entirely
covers the  galaxy. A couple ($A_i,\ {\kappa{}}_{A_i}$),
composed of an observer $A_i$ and a normalization coefficient applied to
  the  spatial  lengths, ${\kappa{}}_{A_i}$, is associated with each open set ${U}_{A_i}$. 
The observers $A_i$ are assumed to be at rest  relatively to each
other, as follows:

\begin{equation}
{\boldsymbol{v}(A_i)}_{A_j}=\boldsymbol{0}
\end{equation}

Then, the  collection $S_{\left\{A_i\right\}}$ composes a reference frame $R_A$. The observers $A_i$ do not
use  the same unit of length along $R_A$, but they are at rest 
relatively to each other as imposed by eq. (28).  Next, for any point $P$ in the
set $U_{A_i}$, the following relationship is used:

\begin{equation}
{\boldsymbol{v}(P)}_{A_i}={\kappa{}}_{A_i}{\dot{\boldsymbol{\sigma{}}}(P)}_{A_i}
\end{equation}

Second, the motion of any galaxy as a whole has to be treated as if this one was a point confounded
with its barycenter. Any  galaxy evolves in a cluster of galaxies. A second collection $S_{\left\{B_l\right\}}$ of open sets $U_{B_l}$ is considered and once again covers  the entire galaxy cluster. The observers $B_l$ are assumed to be
 at rest relatively to each other, as follows:

\begin{equation}
{\boldsymbol{v}(B_l)}_{B_m}=\boldsymbol{0}
\end{equation}

Then, the collection $S_{B_l}$ constitutes  a second reference  frame $R_B$.  A couple ($B_l$, ${\kappa{}}_{B_l}$),
composed of an observer $B_l$ and a normalization coefficient applied to the
 spatial lengths, ${\kappa{}}_{B_l}$, is associated to each open set $U_{B_l}$.
For any  couple of sets $U_{A_i}$ and $U_{A_j}$ located in a set ${U}_{B_l}$, we have the following relationships:

\begin{equation}
{\dot{\boldsymbol{\sigma{}}}(A_i)}_{B_l}={\dot{\boldsymbol{\sigma{}}}(A_j)}_{B_l}
\end{equation}

and

\begin{equation}
{\boldsymbol{v}(A_i)}_{B_l}={\kappa{}}_{B_l}{\dot{\boldsymbol{\sigma{}}}(A_i)}_{B_l}
\end{equation}

\begin{center}
\includegraphics[height=130pt, width=310pt]{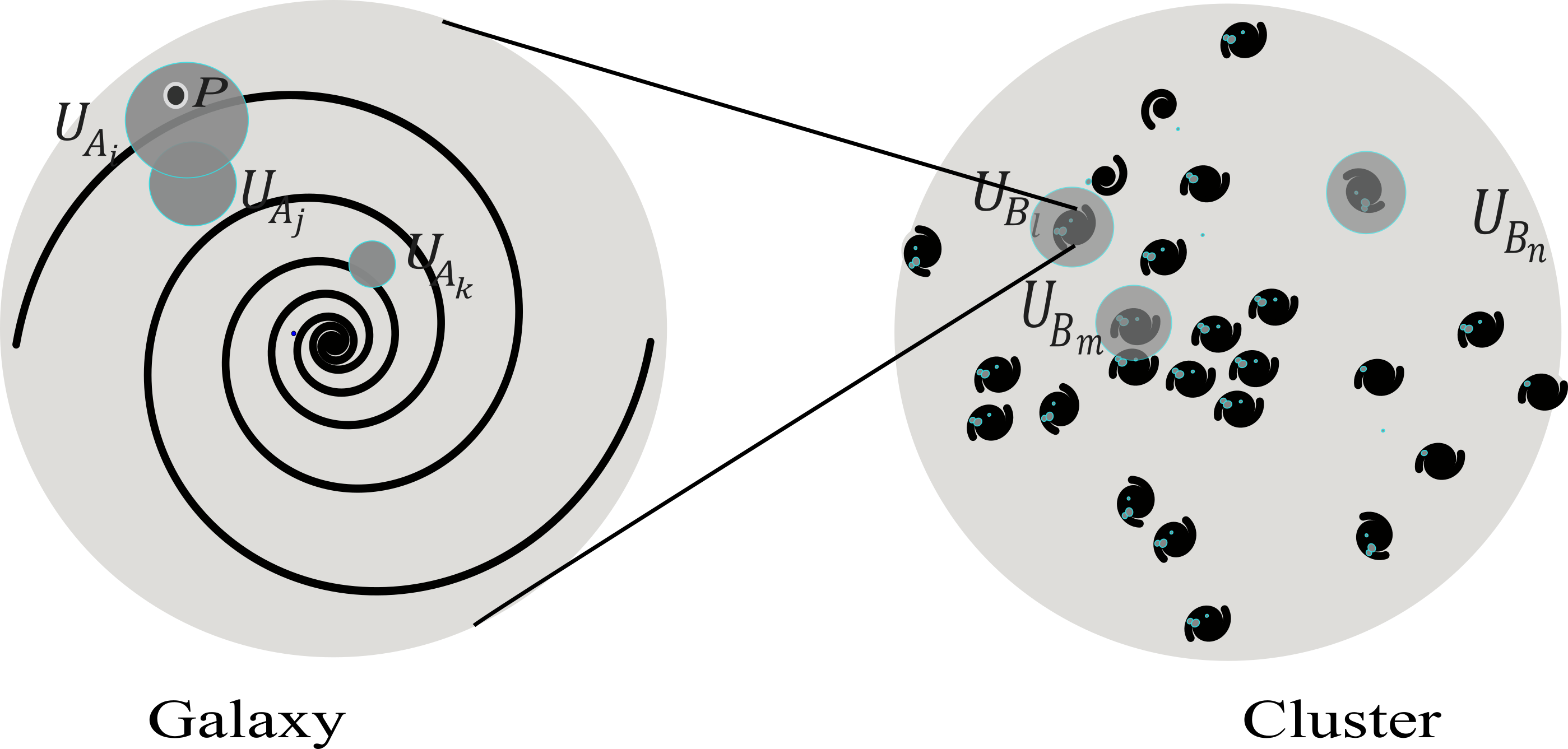}

Figure 9
\end{center}

\noindent 
The law of addition of velocities is applied, as follows:

\begin{equation}
{\boldsymbol{v}(P)}_{B_l}={\boldsymbol{v}(P)}_{A_i}
+{\boldsymbol{v}(A_i)}_{B_l}
\end{equation}

with the following: ${\boldsymbol{v}(P)}_{A_i}={\kappa{}}_{A_i}{\dot{\boldsymbol{\sigma{}}}(P)}_{A_i}$, ${\boldsymbol{v}(A_i)}_{B_l}={\kappa{}}_{B_l}{\dot{\boldsymbol{\sigma{}}}(A_i)}_{B_l}$.

Based on these relationships,  the expression (25) 
can now  be properly  rewritten ($d\boldsymbol{f}_i$ represents the internal forces, $d\boldsymbol{f}_{inertial}$, $d\boldsymbol{f}_e$ are the
inertial forces and  the external forces, respectively, applied
on the mass $dm$), as follows:

\begin{equation}
dm\frac{d}{dt}{(\kappa{}}_{A_i}\dot{\boldsymbol{\sigma{}}}\left(P\right)_{A_i})=d\boldsymbol{f}_i+d\boldsymbol{f}_{inertial}+d\boldsymbol{f}_e
\end{equation}

This equation  treats the case of the internal motions within the galaxy  (for instance, the
rotation). We can  integrate eq. 34 over the volume $\Omega{}$ of the galaxy, as follows:

\begin{equation}
\int_{\Omega{}}dm\frac{d}{dt}({\kappa{}}_{A_i}{\dot{\boldsymbol{\sigma{}}}\left(P\right)}_{A_i})=\int_{\Omega{}}d\boldsymbol{f}_i+ \int_{\Omega{}}d\boldsymbol{f}_{inertial}+\int_{\Omega{}}d\boldsymbol{f}_e=\boldsymbol{F}_i+\boldsymbol{F}_{inertial}+\boldsymbol{F}_e
\end{equation}

For an isolated ($\boldsymbol{F}_e=\boldsymbol{0}$, $\boldsymbol{F}_{inertial}=\boldsymbol{0}$) and symmetric ($\boldsymbol{F}_{i\ a sym}=0$) galaxy, the following relationship is obtained:

\begin{equation}
\int_{\Omega{}}{dm\
\kappa{}}_{A_i}{\dot{\boldsymbol{\sigma{}}}\left(P\right)}_{A_i}=\boldsymbol{K}
\end{equation}

where the constant $\boldsymbol{K}$ can eventually considered nul.

The dynamic equation for a symmetric  galaxy studied as a whole and covered by a set $B_l$ is as follows: \footnote{For
an isolated and  asymmetric galaxy for which $\boldsymbol{F}_{i\ asym}\neq \boldsymbol{0}$, this equation becomes the following:

\begin{equation}
M\frac{d}{dt}({\kappa{}}_{B_l}{\dot{\boldsymbol{\sigma{}}}(A_i)}_{B_l})
=\boldsymbol{F}_{i\ asym}
\end{equation}

\noindent 
In this case, the galaxy can be auto-accelerated.}

\begin{equation}
M\frac{d}{dt}({\kappa{}}_{B_l}{\dot{\boldsymbol{\sigma{}}}(A_i)}_{B_l})
=\boldsymbol{F}_e
\end{equation}

where $M$ is the total  mass  of the galaxy ($M=\int_{\Omega{}}dm)$). The equation (38) can be used to treat
the motion of a galaxy (viewed as a point here) in a galaxy cluster.   The dynamic equation (34) relative to any point $P$ of mass $m$ in the galaxy becomes:\footnote{Let $P$ contained in the set  $A_i$ and $P’$ contained in the
set ${ A}_{i'}$  The internal forces are calculated as follows:

\begin{equation}
\boldsymbol{f}_{P'\longrightarrow{}P}=-G\frac{{\kappa{}}_{A_i}\
\boldsymbol{P'P}}{{{\kappa{}}_{A_i}^3}{{P'P}}^3}  \ \ \ \  \ 
\boldsymbol{f}_{P\longrightarrow{}P'}=-G\frac{{\kappa{}}_{A_{i'}}\boldsymbol{
PP'}}{{{\kappa{}}_{A_{i'}}^3}{PP'}^3}
\end{equation}

The force needs to be  measured in situ  (i.e. respectively at $P$ for $\boldsymbol{f}_{P'\longrightarrow{}P}$ and at $P'$ for $\boldsymbol{f}_{P\longrightarrow{}P'})$ and, in this case, the measured force is a well-defined quantity and considered observer-independent.}

\begin{equation}
m(\frac{d}{dt}{\kappa{}}_{A_i}{\dot{\boldsymbol{\sigma{}}}\left(P\right)}_{A_i})=\boldsymbol{f}_i+(\boldsymbol{f}_e-\frac{m}{M}\boldsymbol{F}_e)
\end{equation}

The force $\boldsymbol{f}_e-\frac{m}{M}\boldsymbol{F}_e$ is the tidal force produced by an outer mass. Equation (40) can be 
used to treat the internal motions in a galaxy.

\end{document}